\documentclass[%
 aip,
 amsmath,amssymb,
preprint,%
]{revtex4-1}

\usepackage{appendix}

\usepackage{graphicx}
\usepackage{dcolumn}
\usepackage{bm}

\usepackage[utf8]{inputenc}
\usepackage[T1]{fontenc}
\usepackage{mathptmx}
\usepackage{etoolbox}
\usepackage{float}
\usepackage{color}

\makeatletter
\def\@email#1#2{%
 \endgroup
 \patchcmd{\titleblock@produce}
  {\frontmatter@RRAPformat}
  {\frontmatter@RRAPformat{\produce@RRAP{*#1\href{mailto:#2}{#2}}}\frontmatter@RRAPformat}
  {}{}
}%
\makeatother
\begin{document}

\preprint{AIP/123-QED}

\title{Vector Representation of Exact Soliton Dynamics in Multi-component Nonlinear Schrödinger Systems}
\author{M. Foucher}
 \affiliation{LyRIDS, ECE Engineering School, OMNES Education, 10 rue Sextius Michel, 75015, Paris, France}
\author{L. Delisle}%
\affiliation{LyRIDS, ECE Engineering School, OMNES Education, 10 rue Sextius Michel, 75015, Paris, France}%

\author{A. Jaouadi}
 \email{ajaouadi@ece.fr}
\affiliation{LyRIDS, ECE Engineering School, OMNES Education, 10 rue Sextius Michel, 75015, Paris, France}%

\date{\today}

\begin{abstract}
Multicomponent nonlinear Schrödinger equations constitute fundamental models for coherent matter waves in multicomponent Bose--Einstein condensates, spinor quantum fluids, and vector nonlinear optical systems. We develop a vector formulation of the Hirota bilinear formalism for the completely integrable Manakov system that treats the coupled nonlinear Schrödinger equations directly at the vector level rather than through the conventional component-wise decomposition. This framework naturally retains the intrinsic multicomponent representation of the model while providing compact analytical expressions for exact vector soliton solutions. Within this approach, we systematically construct bright, dark, and mixed one-, two-, and three-soliton solutions and show how the underlying vector structure provides a unified description of their nonlinear interactions. In particular, the proposed formalism makes the coupling between the different components explicit while preserving the geometric organization of the vector system throughout the bilinearization procedure. Beyond its analytical simplicity, the framework offers a natural perspective for the study of coherent multicomponent nonlinear excitations and provides a foundation for extending vector Hirota methods to other classes of exact solutions, including rogue waves, periodic waves, and rational solutions.

\end{abstract}

\maketitle

\section{Introduction}

Integrable systems play a central role in mathematical physics because they constitute one of the few classes of nonlinear models for which exact analytical solutions can be obtained. Among their most remarkable manifestations are solitons, localized nonlinear wave packets that propagate without changing shape and preserve their identities after interactions with other solitary waves. Beyond their mathematical significance, solitons are now recognized as fundamental coherent excitations in a broad range of physical systems, including quantum fluids, nonlinear optical media, and matter-wave systems. For nonlinear Schrödinger equations, one generally distinguishes between bright and dark solitons, corresponding, respectively, to attractive and repulsive nonlinear interactions \cite{Becker2008}. Bright solitons appear as localized density peaks evolving on a vanishing background, while dark solitons manifest as localized density depressions embedded in a finite and uniform background. Their interactions exhibit characteristic nonlinear signatures: bright solitons may generate strong interference patterns, whereas dark solitons undergo collision-induced phase shifts that can be described analytically within integrable models\cite{DelisleSym, Celanie,DelisleDark}. Other classes of solitary waves arise in nonlinear evolution equations, such as the Korteweg--de Vries equation, where solitons describe stable wave propagation in shallow-water environments. In modern physics, Schrödinger solitons have found important applications in Bose--Einstein condensates (BECs), nonlinear optics, and even econophysics through the Ivancevic model \cite{Ivancevic,IvanDel}, where nonlinear Schrödinger equations are employed to describe the time evolution of option prices.

Over the years, physicists and mathematicians have developed a variety of analytical techniques to investigate integrability and construct exact soliton solutions. Among the most prominent methods are Lie point symmetries\cite{Olver}, the inverse scattering transform (IST)\cite{Ablowitz}, Darboux and Bäcklund transformations\cite{Rogers}, and Hirota's bilinear formalism\cite{Hirota,Hietarinta}. The latter algebraic approach has proven particularly successful for obtaining exact multi-soliton solutions in closed form. It relies on transforming the original nonlinear system into an equivalent bilinear representation expressed through Hirota differential operators, which will be introduced later in this paper. The resulting bilinear equations are then solved through finite $\epsilon$-expansions, where each solitary wave is represented by an independent exponential mode. A central result of Hirota theory is that the existence of unconstrained three-soliton solutions provides a strong test of complete integrability. Hietarinta systematically classified bilinear equations satisfying Hirota's three-soliton condition and demonstrated that integrable systems admit multi-soliton solutions expressed as finite polynomials of exponential functions \cite{Hietarinta1,Hietarinta2,Hietarinta3,Hietarinta4}. This criterion is closely related to integrability in the sense of the inverse scattering transform.

Since its introduction, the Hirota bilinear formalism has evolved far beyond its original role as a method for constructing multi-soliton solutions. Its flexibility has enabled extensions to a wide variety of integrable systems and nonlinear wave phenomena. The formalism has been successfully adapted to nonlocal integrable equations, including complex modified Korteweg--de Vries equations and coupled nonlinear Schrödinger systems \cite{LiLi,Bai}, thereby expanding the range of physically relevant models accessible to exact analytical treatment. It has also been generalized to supersymmetric integrable systems \cite{Carstea,ZhangLiu,Grammaticos,DelisleMas}, providing a unified framework for the study of nonlinear excitations in multidimensional Grassmann superspaces and supersymmetric extensions of classical models such as the Korteweg--de Vries and Sine--Gordon equations.

Beyond soliton solutions, the Hirota formalism has proven remarkably effective for constructing a broad spectrum of coherent nonlinear structures, including rogue waves, rational solutions, and lump solutions \cite{Ma,Yue}. In particular, rational solutions are closely connected to special polynomial hierarchies arising in integrable systems, such as the Yablonskii--Vorob'ev polynomials \cite{Clarkson}, which play an important role in the construction of similarity solutions of Painlevé-type equations. Such waveforms are of considerable physical interest in nonlinear optics, hydrodynamics, plasma physics, and matter-wave systems, where extreme localization and nonlinear focusing phenomena play a central role. In parallel, connections between Hirota bilinear forms and Bell polynomials have provided powerful algebraic tools for deriving bilinear representations and higher-order nonlinear structures \cite{MaBell,MaTril}, while the development of linear superposition principles and invariant solution subspaces has further enriched the mathematical structure of the method \cite{MaSub}. Collectively, these developments demonstrate that the Hirota formalism is not merely a technique for generating exact solutions but a versatile framework for exploring the rich dynamics of integrable nonlinear systems and their applications to a broad range of physical phenomena.

The Hirota bilinear formalism has also been extensively applied to coupled integrable systems, including coupled modified Korteweg--de Vries equations \cite{DelisleOCNMP,Hirota1,Iwao} and multi-component nonlinear Schrödinger equations \cite{Ohta,Jiang,Vijaya,Radha,Rad,Kanna1,MaoZhao}. Such models play a central role in the theory of nonlinear waves because they describe the interaction of several coherent fields and support a rich variety of vector soliton states. In particular, multi-component nonlinear Schrödinger equations provide an effective theoretical framework for the study of vector matter waves in multi-component Bose--Einstein condensates, spinor condensates, and nonlinear optical media supporting polarization dynamics \cite{Trippenbach2000,Kevrekidis2016,Mollenauer,Hasegawa}. These systems exhibit phenomena that have no counterpart in scalar models, including vector solitons, dark--bright complexes, polarization effects, and inter-component nonlinear interactions.

Within the conventional Hirota approach, a separate bilinear transformation is introduced for each component of the system, leading to a hierarchy of coupled bilinear equations. Although this procedure successfully generates exact multi-soliton solutions, the resulting representation often obscures the intrinsic vector structure, symmetry properties, and collective nature of the underlying model. As the number of components increases, the component-wise formulation becomes increasingly cumbersome and tends to hide the geometric relations responsible for the nonlinear coupling between the different fields. These observations naturally motivate the search for formulations that preserve the vector character of the original system throughout the bilinearization procedure.

Recently, a vector extension of the Hirota bilinear formalism was introduced, in which coupled nonlinear systems are treated directly at the vector level rather than through a component-wise decomposition \cite{DelisleOCNMP}. Applied to a coupled system of modified Korteweg--de Vries equations, this framework preserved the underlying vector structure of the model and made the nonlinear interactions between the different components explicit through a coupling-matrix representation. Beyond its mathematical elegance, such an approach is particularly
attractive for multi-component nonlinear systems because it preserves
the natural symmetry properties of the model and provides a direct
description of collective vector excitations encountered in
multi-component BECs and related matter-wave
systems \cite{LiuPu}.

In the present work, we apply this vector Hirota framework to the completely integrable multi-component nonlinear Schrödinger equation of Manakov type\cite{Manakov}. Our objective is not to derive previously unknown soliton solutions, but rather to develop a structure-preserving vector formulation that retains the intrinsic representation structure of the model throughout the bilinearization procedure. Within this framework, we systematically construct bright, dark, and mixed one-, two-, and three-soliton solutions directly at the vector level. The resulting representation provides compact analytical expressions while making the nonlinear interactions between the different components explicit. More generally, it offers a unified description of multi-component coherent structures and highlights the geometric and collective aspects of vector soliton dynamics. Such excitations are of broad interest in the study of multi-component BECs, spinor matter waves, and vector nonlinear optical systems, where internal degrees of freedom play a fundamental role.

The paper is organized as follows. In Sec.~II, we introduce the vector formulation of the multi-component nonlinear Schrödinger system and discuss, in particular, its symmetry properties. In Sec.~III, we establish the corresponding vector Hirota bilinear representation and show how the structure of the ground state naturally distinguishes between bright, dark, and mixed vector soliton sectors. In Sec.~IV, we construct explicit one-, two-, and three-soliton solutions and analyze their dynamical properties. Finally, Sec.~V summarizes the main results and outlines possible extensions of the vector Hirota formalism toward rogue waves, rational solutions, breather states, and other classes of coherent nonlinear excitations.

\section{The model}

As a prototypical example of an integrable multi-component nonlinear system, we consider the integrable $N$-component nonlinear Schrödinger equation of Manakov type

\begin{equation}
i \Psi_t+\frac12 \Psi_{xx}+\gamma (\Psi^{\dagger}\Psi)\Psi=0,
\label{Master}
\end{equation}
where $\gamma$ is a nonzero real constant characterizing the strength of the nonlinear interaction and $\Psi$ is a complex-valued vector field with $N$ components,

\begin{equation}
\Psi^{t}=
\left(
\begin{array}{cccc}
\psi_1, & \psi_2, & \cdots, & \psi_N
\end{array}
\right),
\label{vectorPsi}
\end{equation}
whose components $\psi_j=\psi_j(x,t)$ depend on the spatial coordinate $x$ and time $t$. Throughout this work, the superscripts  $t$ and $\dagger$ denote, respectively, vector transposition and Hermitian conjugation.

Equation (\ref{Master}) constitutes a natural generalization of the celebrated Manakov system \cite{Manakov}, recovered in the particular case $N=2$. Owing to its complete integrability, this model has played a fundamental role in the study of vector nonlinear waves and multi-component coherent structures. In physics, it provides an effective description of a variety of systems, including multi-component Bose--Einstein condensates, nonlinear optical media, and other coupled-wave configurations\cite{Trippenbach2000,Kevrekidis2016,Mollenauer,Hasegawa,Zhang}. The model supports a rich spectrum of coherent excitations, including bright, dark, and mixed vector solitons, whose dynamics are governed by the interplay between nonlinearity and dispersion.

A key property of Eq.~(\ref{Master}) is its invariance under global unitary transformations. More precisely, Eq.~(\ref{Master}) is invariant under the action of the unitary group $U(N)$, where $U(N)$ denotes the set of complex-valued matrices satisfying

\[
U^{\dagger}U=UU^{\dagger}=\mathbb{I}_N ,
\]
with $\mathbb{I}_N$ the $N\times N$ identity matrix. Indeed, if $\Psi$ is a solution of Eq.~(\ref{Master}), then $\Psi' = U\Psi$, is also a solution for any $U\in U(N)$. This symmetry plays a central role in the present work because it naturally motivates a formulation directly expressed at the vector level and allows the intrinsic structure of the multi-component system to be preserved throughout the bilinearization procedure.

To clarify future notations, let  $\{e_1,e_2,\cdots,e_N\}$ denote the canonical basis of $\mathbb{R}^N$, where $e_j$ corresponds to the $j^{\mathrm{th}}$ column of the identity matrix $\mathbb{I}_N$. These vectors satisfy the standard orthonormality relations 
$e_m^{\dagger}e_n=\delta_{mn}$, where $\delta_{mn}$ denotes the Kronecker delta.

Hirota constructions for multi-component nonlinear Schrödinger equations proceed through a component-wise decomposition of the vector field, introducing a separate bilinear transformation for each component\cite{Ohta,Jiang,Vijaya,Radha,Rad,Kanna1}. While this approach generates exact vector soliton solutions, it tends to obscure the symmetry structure of the original model and the geometric nonlinear relationships between its different components. In contrast, the objective of the present work is to develop a bilinear formulation directly at the vector level. The resulting framework provides a compact and transparent description of vector soliton states and their nonlinear interactions.

\section{Vector formulation of the Hirota Bilinear Formalism}

In this section, we use the recent vector adaptation of the Hirota bilinear formalism\cite{DelisleOCNMP} to give a vector bilinear representation of the coupled system of nonlinear Schrödinger equations (\ref{Master}). The Hirota equations are given directly at the vector level instead of the usual component-wise approach.

\subsection{Ground state solution}

In order to construct a vector Hirota bilinear representation of Eq.~(\ref{Master}), we first identify a uniform ground state solution. This configuration plays a central role in the present construction, as it determines the background on which nonlinear excitations evolve and naturally leads to the classification of bright, dark, and mixed vector soliton families. Owing to the $U(N)$ symmetry of the model, all constant-density states connected by a global unitary transformation are physically equivalent. It is therefore sufficient to select the symmetric representative
\begin{equation}
    \Psi_0(x,t)=\frac{e^{i(\kappa x+\omega t)}}{\sqrt{N}}\sum_{j=1}^N
e_j,\label{GroundState}\end{equation}
where $\kappa$ and $\omega$
 are real constants to be determined. We have $\vert \Psi_0\vert^2=\Psi_0^{\dagger}\Psi_0=1$. Introducing the ansatz (\ref{GroundState}) in Eq.(\ref{Master}), we find
 \begin{equation}
     \omega=-\frac{\kappa^2}{2}+\gamma.\label{dispersion}
 \end{equation}

 The family of background states (\ref{GroundState}) therefore forms a homogeneous finite-density manifold related by $U(N)$ transformations. This symmetry will be preserved throughout the vector bilinearization procedure and will play a central role in the construction of vector soliton solutions.

\subsection{Vector Hirota bilinear representation}

Motivated by the structure of the background solution (\ref{GroundState}), we introduce the following vector Hirota transformation:
\begin{equation}
    \Psi=\frac{e^{i(\kappa x+\omega t)}}{\sqrt{N}}\frac{F}{G},\label{HirotaAnzatz}
\end{equation}
 where $\omega$ satisfies the dispersion relation (\ref{dispersion}), $F=F(x,t)$ is a complex-valued $N-$components vector and $G=G(x,t)$ is a real-valued function. We write
 \begin{equation}
     F(x,t)=\sum_{k=1}^N f_k(x,t)e_k,
 \end{equation}
 where $f_k$ are the complex-valued components of the vector $F$. Introducing the ansatz (\ref{HirotaAnzatz}) in Eq.(\ref{Master}) gives the vector Hirota bilinear representation
 \begin{eqnarray}
\left(i\mathcal{D}_t+i\kappa \mathcal{D}_x+\frac12\mathcal{D}_x^2-\gamma-\beta\right)(F\cdot G)&=&0,\label{HirotaEq1}\\
\frac12\mathcal{D}_x^2(G\cdot G)-\frac{\gamma}{N}\vert F\vert^2-\beta \,G^2&=&0\label{HirotaEq2},
 \end{eqnarray}
where $\beta$ is a constant parameter that plays a fundamental role in the classification of vector soliton families. The vector Hirota formalism does not merely provide an alternative representation of the bilinear equations; it preserves the $U(N)$ symmetry of the original multi-component model throughout the bilinearization procedure.
Unlike the conventional component-wise Hirota formalism, Eq.~(\ref{HirotaEq1}) is formulated directly at the vector level. The bilinear operator therefore acts on the vector field $F$ as a whole while preserving the intrinsic multi-component structure of the system. The vector equation (\ref{HirotaEq1}) is equivalent to the following set of $N$ scalar bilinear equations:
 \begin{equation}
    \sum_{\mu=1}^N \left(i\mathcal{D}_t+i\kappa \mathcal{D}_x+\frac12\mathcal{D}_x^2-\gamma-\beta\right)(f_{\mu}\cdot G)e_{\mu}=0.
 \end{equation}
In our recent work, we have shown that the Hirota vector formulation is basis-independent, so that the canonical basis of $\mathbb{R}^N$ may be used for convenience. Eq.(\ref{HirotaEq2}) is a classical scalar equation. The operator $\mathcal{D}$ is known as the Hirota derivative and is defined as
\begin{equation}
    \mathcal{D}_z^n(\mathcal{F}\cdot \mathcal{G})=(\partial_{z_1}-\partial_{z_2})^n\mathcal{F}(z_1)\mathcal{G}(z_2)\vert _{z_1=z_2=z}
\end{equation}
for $\mathcal{F}$ and $\mathcal{G}$ scalar functions. A notable consequence of Eq.~(\ref{HirotaEq2}) is that the total density $\vert\Psi\vert^2$ admits a compact closed-form representation: 
\begin{equation}
    \vert\Psi\vert^2=\sum_{k=1}^N\vert \psi_k\vert^2=\frac{1}{NG^2}\sum_{k=1}^N\vert f_k\vert^2=\frac{1}{\gamma}\partial_x^2\ln G-\frac{\beta}{\gamma}.\label{DensityVec}
\end{equation}

In order to construct multi-soliton solutions from the Hirota bilinear system (\ref{HirotaEq1})-(\ref{HirotaEq2}), we need to extract the constant ground state couple $(F_0,G_0)$ for $F_0\in \mathbb{C}^N$ and $G_0\in\mathbb{R}$. We may assume that $G_0=1$ and, using the $U(N)$ invariance of Eq.(\ref{Master}), we take $F_0\in \mathbb{R}^N$. The ground state couple satisfy :
\begin{equation}
    -\gamma\, F_0=\beta\, F_0\quad \mbox{and}\quad
-\gamma \vert F_0\vert^2=\beta N.\end{equation}
The admissible ground-state non-uniform background obtained from the bilinear system naturally separates the solution space into distinct classes of nonlinear excitations. Two fundamental backgrounds emerge:
\begin{enumerate}
    \item $F_0=0$ and $\beta=0$. In this case, the resulting nonlinear excitations correspond to localized bright vector solitons propagating on a uniform zero  background.
    \item $\vert F_0\vert^2=N$ and $\beta=-\gamma$. One choice for $F_0$, using $U(N)-$invariance, is the uniformly distributed vector
    \begin{equation}
        F_0=\sum_{\mu=1}^Ne_{\mu}.
        \label{BackDD}
    \end{equation}
    In this case, the solutions manifest as intensity depressions from a maximal density background. More generally, non-uniform choices of $F_0$ may be considered, where only a subset of the vector components carries a non-zero finite background density. As will be shown later, these configurations naturally generate mixed bright-dark vector soliton states from orthogonal spaces spanning the bright and dark contributions, respectively, of the vector soliton construction.
\end{enumerate}

 \section{Soliton solutions}

Having established the vector Hirota bilinear representation of Eq.~(\ref{Master}), we now construct explicit vector multi-soliton solutions of the coupled nonlinear Schrödinger system. The Hirota method generates nonlinear coherent structures as finite perturbations of a chosen ground-state configuration. Consequently, the nature of the resulting excitations is directly determined by the background pair $(F_0,G_0)$ introduced in the previous section. In particular, uniformly vanishing backgrounds give rise to bright vector solitons, uniformly finite-density backgrounds generate dark vector solitons, while partially populated backgrounds lead to mixed bright-dark vector states.

To obtain these solutions, we solve the bilinear system (\ref{HirotaEq1})--(\ref{HirotaEq2}) through Hirota $\epsilon$-expansions around the ground-state pair $(F_0,G_0)$:

\begin{equation}
F=F_0+\epsilon F_1+\epsilon^2F_2+\cdots\quad \mbox{and} \quad G=1+\epsilon G_1+\epsilon^2 G_2+\cdots,
\label{vectorexpan}
\end{equation}
where $F_k$ are complex-valued $N$-component vector functions and $G_k$ are real-valued scalar functions. Within Hirota theory, exact multi-soliton solutions arise from finite truncations of these expansions. Each independent exponential mode introduced in the construction corresponds to an elementary vector solitary wave, while nonlinear interactions between solitons emerge naturally through higher-order terms generated by the bilinear equations. In the following, we construct the vector three-soliton solution for all types, highlighting the integrability of the system in the Hirota sense.

\subsection{Soliton solutions in a uniform zero background}

We first consider the sector associated with the vanishing ground-state configuration $F_0=0$ and $\beta=0$. In this regime, the nonlinear excitations evolve on a zero-density background and therefore correspond to bright vector solitons. Physically, these solutions describe localized wave packets whose density vanishes asymptotically far from the soliton core. The absence of a finite background considerably simplifies the bilinear structure and leads to the reduced system

\begin{eqnarray}
\left(i\mathcal{D}_t+i\kappa \mathcal{D}_x+\frac12\mathcal{D}_x^2-\gamma\right)(F\cdot G)&=&0,\label{HirotaBBEq1} \\
\frac12\mathcal{D}_x^2(G\cdot G)-\frac{\gamma}{N}\vert F\vert^2&=&0,\label{HirotaBBEq2}
\end{eqnarray}
which constitutes the starting point for the construction of bright vector multi-soliton states.

To obtain vector $M$-bright-soliton solutions, we assume that the Hirota expansions (\ref{vectorexpan}) truncate after a finite number of terms according to

\begin{equation}
F=\sum_{k=1}^M\epsilon^{2k-1}F_{2k-1}\quad \mbox{and}\quad G=1+\sum_{k=1}^M\epsilon^{2k}G_{2k}.
\label{expanBB}
\end{equation}

Within Hirota theory, the integer $M$ determines the number of elementary solitary waves that compose the solution. The finite truncation of the series is a hallmark of integrability, as Hirota integrability imposes a finite polynomial construction of exponential functions,  and guarantees the existence of exact multi-soliton solutions. Substituting the expansions (\ref{expanBB}) into the bilinear system (\ref{HirotaBBEq1})--(\ref{HirotaBBEq2}) and equating each power of $\epsilon$ to zero yields a hierarchy of bilinear equations. These equations are solved using $M$ independent exponential modes, each representing an individual vector one-soliton, while higher-order terms encode their nonlinear interactions.

\subsubsection{The vector one-soliton solution}

The vector one-soliton solution is obtained taking $M=1$ in the finite $\epsilon-$expansions (\ref{expanBB}). We get the system of equations:
\begin{eqnarray}
    \left(i\partial_t+i\kappa\partial_x+\frac12\partial_x^2-\gamma\right)F_1&=&0,\label{EBB1}\\
    G_{2,xx}-\frac{\gamma}{N}\vert F_1\vert^2&=&0,\label{EBB2}\\
    \left(i\mathcal{D}_t+i\kappa \mathcal{D}_x+\frac12\mathcal{D}_x^2-\gamma\right)(F_1\cdot G_2)&=&0,\\
    \mathcal{D}_x^2(G_2\cdot G_2)&=&0.
\end{eqnarray}
The above system of equations is solved using the forms
\begin{equation}
    F_1=v\, e^{\Xi}\quad \mbox{and}\quad G_2=a\, e^{\Xi+\Xi^*},
\end{equation}
where $v\in\mathbb{C}^N$ is a constant vector, $a\in\mathbb{R}$ and $\Xi=\Theta x+\Omega t+\Phi$. We get
\begin{equation}
    \Omega=-\kappa \Theta+\frac{i}{2}\Theta^2-i\gamma \quad\mbox{and}\quad a=\frac{\gamma\vert v\vert^2}{N(\Theta+\Theta^*)^2}.
\end{equation}
As a special example, we take $\epsilon^2=a^{-1}$ and obtain the known lump-type exact solution
\begin{equation}
    \vert\Psi\vert^2=\frac{(\Theta+\Theta^*)^2}{4\gamma}\mbox{sech}^2\left(\frac{\Xi+\Xi^*}{2}\right).
\end{equation}

The vector one-soliton solution is completely characterized by the parameters $\Theta$ and $v$. The complex parameter $\Theta$ governs the localization and propagation properties of the solitary wave through the phase variable $\Xi$, while $v$ specifies its multi-component structure.

\subsubsection{The vector two-soliton solution}

The vector two-soliton solution is obtained taking $M=2$ in (\ref{expanBB}) and setting
\begin{equation}
    F_1=v_1\, e^{\Xi_1}+v_2\, e^{\Xi_2},
\end{equation}
where $\Xi_j=\Theta_j \, x+\Omega_j\, t+\Phi_j$ and $v_j\in\mathbb{C}^N$ are constant vectors. The function $F_1$ satisfies Eq.(\ref{EBB1}) and, by linearity, we get
\begin{equation}
    \Omega_j=-\kappa\Theta_j+\frac{i}{2}\Theta_j^2-i\gamma.\label{disperseBB}
\end{equation}
The function $G_2$ satisfies Eq.(\ref{EBB2}) and following the form of $F_1$, we make the ansatz:
\begin{equation}
    G_2=a_{1}^1e^{\Xi_1+\Xi_1^*}+a_{1}^2e^{\Xi_1+\Xi_2^*}+a_{2}^1e^{\Xi_2+\Xi_1^*}+a_{2}^2e^{\Xi_2+\Xi_2^*}.
\end{equation}
We obtain
\begin{equation}
    a_{p}^q=\frac{\gamma}{N}\frac{v_q^{\dagger}v_p}{(\Theta_p+\Theta_q^*)^2}.\label{apq}
\end{equation}
We have $(a_q^p)^{\dagger}=a_p^q$. For $F_3$, we have
\begin{equation}
    F_3=v_{12}^1\, e^{\Xi_1+\Xi_2+\Xi_1^*}+v_{12}^2\,e^{\Xi_1+\Xi_2+\Xi_2^*},\quad v_{12}^p=A_{12}\left(\frac{a_2^p}{\Theta_1+\Theta_p^*}v_1-\frac{a_1^p}{\Theta_2+\Theta_p^*}v_2\right),
\end{equation}
where $A_{12}=\Theta_1-\Theta_2$ is a binary measurement of nonlinear interactions between solitons. For two distinct solitary waves, we have $A_{12}\neq0$. For $G_4$, we obtain
\begin{equation}
    G_4=a_{12}^{12}e^{\Xi_1+\Xi_2+\Xi_1^*+\Xi_2^*},\quad a_{12}^{12}=\vert A_{12}\vert^2\left(\frac{a_1^1a^2_2}{(\Theta_1+\Theta_2^*)(\Theta_2+\Theta_1^*)}-\frac{a_1^2a_2^1}{(\Theta_1+\Theta_1^*)(\Theta_2+\Theta_2^*)}\right).
\end{equation}

The structure of the two-soliton solution naturally reflects the nonlinear interaction between two elementary vector solitons. The vectors $v_1$ and $v_2$ characterize the individual solitary waves, while the vector $v_{12}^p$ is generated through their nonlinear coupling in the bilinear construction. Similarly, the coefficients $a_{12}^{12}$ and $A_{12}$ arise from interaction terms and account for the corrections required to construct an exact two-soliton solution. This hierarchy of vector and scalar coefficients is a characteristic feature of the Hirota formalism and will naturally extend to the three-soliton case discussed below.

\begin{figure*}[h]
    \centering
    \includegraphics[width=0.9\textwidth]{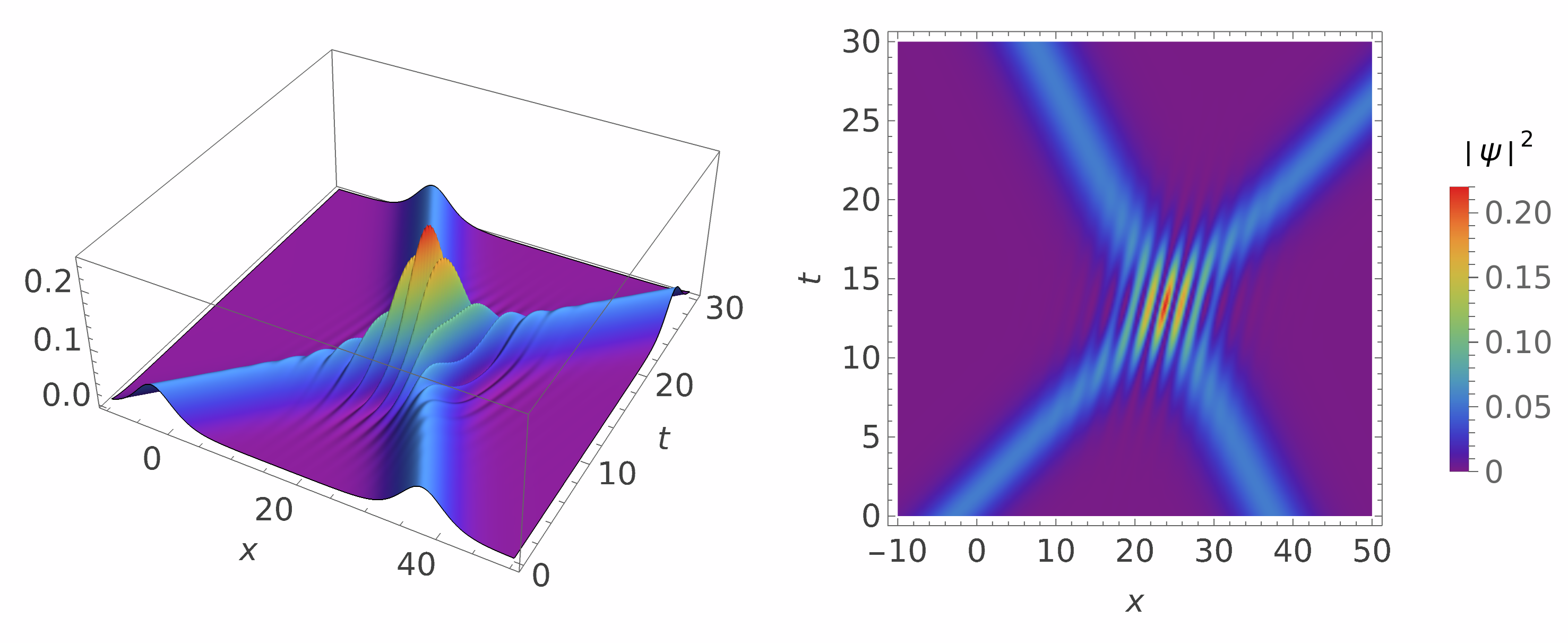}
    \caption{The density function $\vert\Psi(x,t)\vert^2$ given in Eq.(\ref{DensityVec}) for $\gamma=\kappa=\epsilon=1$, $N=2$, $\Theta_1=0.25-2i$, $\Theta_2=0.25+i$, $\Phi_1=-10$, $\Phi_2=0$, $v_1^t=(1\,,\,\,i)
    $ and $v_2^t=(-i\,\,,\,1)
    $.}
    \label{bright2SoluionVec}
\end{figure*}

In Figure~\ref{bright2SoluionVec}, we present the density function $\vert\Psi\vert^2$ for the system (\ref{Master}) with $N=2$. 
corresponding to the classical Manakov model. We choose $\gamma=\kappa=\epsilon=1$, $\Theta_1=0.25-2i$, $\Theta_2=0.25+i$, $\Phi_1=-10$, $\Phi_2=0$, $v_1^t=(1,i)$ and $v_2^t=(-i,1)$. The dynamics are characteristic of a bright two-soliton solution. At early and late times, the two localized excitations propagate as well-separated solitary waves. As they approach each other, a pronounced interference pattern develops in the interaction region, reaching its maximum intensity near the collision time $t\approx13$. Following the interaction, the two solitons recover their localized profiles and continue their propagation without observable distortion, illustrating the elastic nature of soliton collisions in completely integrable systems.

\subsubsection{The vector three-soliton solution}

The construction of the vector three-soliton solution constitutes a nontrivial consistency test of the proposed vector Hirota formalism. In Hirota theory, the existence of an exact three-soliton solution without additional constraints is generally regarded as a hallmark of complete integrability. Consequently, the present result demonstrates that the vector bilinear representation preserves the integrable structure of the multi-component nonlinear Schrödinger system while maintaining its intrinsic vector character.

To construct the vector three-soliton solution, we take $M=3$ in Eq.(\ref{expanBB}) and set
\begin{equation}
    F_1=v_1\, e^{\Xi_1}+v_2\, e^{\Xi_2}+v_3\, e^{\Xi_3},\label{3SolF1BB}
\end{equation}
where $\Xi_j=\Theta_j \, x+\Omega_j\, t+\Phi_j$ and $v_j\in\mathbb{C}^N$ are constant vectors. The function $F_1$ satisfies Eq.(\ref{EBB1}) and, by linearity, we get the dispersion relations (\ref{disperseBB}). To keep the main text concise, the expressions of the other components are provided in the appendix~\ref{appendixA1}.

The hierarchical structure of the solution naturally extends that of the two-soliton case. In addition to the elementary vectors $v_1$, $v_2$, and $v_3$ associated with the individual solitons, the solution contains interaction-generated contributions involving pairs and triplets of solitons. These terms account for the nonlinear couplings that occur when several vector solitary waves coexist and interact within the same integrable system.

\begin{figure*}[h]
    \centering
    \includegraphics[width=0.85\textwidth]{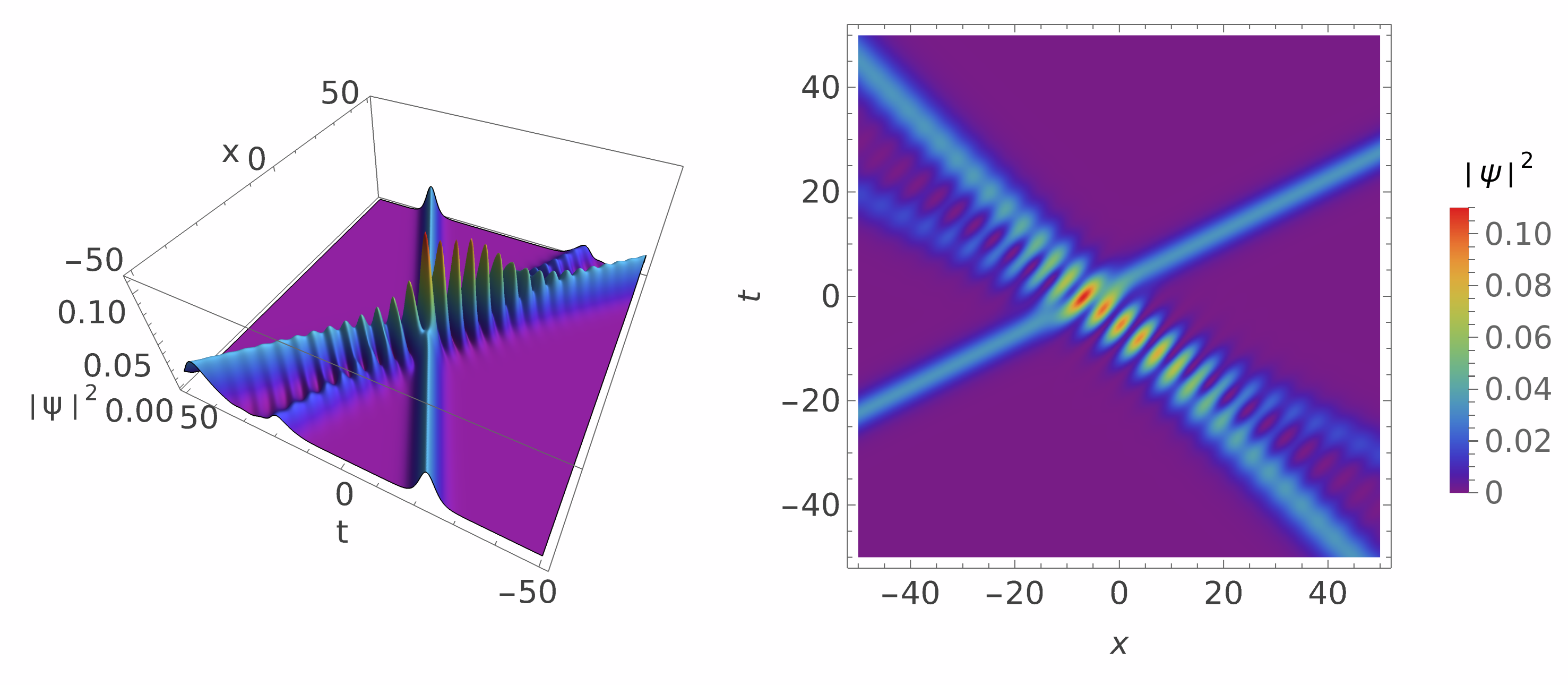}
    \caption{The density function $\vert\Psi(x,t)\vert^2$ given in Eq.(\ref{DensityVec}) for $\gamma=\kappa=\epsilon=1$ and $N=2$. The other parameters are chosen as: $\Theta_1=0.1875-2i$, $\Theta_2=0.125-i$, $\Theta_3=0.1875+i$, $\Phi_1=\Phi_2=\Phi_3=0$, $v_1^t=v_2^t=
        (1\,\,,\,\,i)
    $ and $v_3^t=(
        i\,\,,\,\,1
    )$}
    \label{bright3SoluionVec}
\end{figure*}

In Figure~\ref{bright3SoluionVec}, we present the density function $\vert\Psi\vert^2$ for system (\ref{Master}) with $N=2$, corresponding to the classical Manakov model. We choose $\gamma=\kappa=\epsilon=1$, $\Theta_1=0.1875-2i$, $\Theta_2=0.125-i$, $\Theta_3=0.1875+i$, $\Phi_1=\Phi_2=\Phi_3=0$, $v_1^t=v_2^t=(1, i)$, and $v_3^t=(i,1)$. The resulting dynamics are characteristic of a bright three-soliton solution. The three localized excitations propagate with different velocities and undergo a sequence of nonlinear interactions as they approach each other. Compared with the two-soliton case, the interaction pattern is considerably richer, exhibiting multiple collision events and regions where the influence of several solitons overlaps simultaneously. Pronounced interference structures and temporary density enhancements appear during these interactions, while the individual solitons remain clearly identifiable. After each collision, the localized wave packets recover their original profiles and continue their propagation without observable distortion. This robustness under repeated interactions is a hallmark of multi-soliton dynamics in completely integrable systems and illustrates the ability of the vector Hirota formalism to describe increasingly complex nonlinear processes while preserving the exact solvability of the model.


\subsection{Soliton solution in a uniform maximal density background}

In contrast to the bright-soliton sector discussed previously, the present class of solutions evolves on a uniform finite-density background determined by the non-vanishing ground state $F_0$. As a result, nonlinear excitations manifest themselves as localized density depressions rather than localized density peaks. The finite background fundamentally modifies the interaction dynamics and leads to characteristic dark-soliton signatures, including collision-induced phase shifts and trajectory displacements.

This case corresponds to $\vert F_0\vert^2=N$ and $\beta=-\gamma$. In this formulation, the system of bilinear equations (\ref{HirotaEq1}) and (\ref{HirotaEq2}) reduces to
 \begin{eqnarray}
\left(i\mathcal{D}_t+i\kappa \mathcal{D}_x+\frac12\mathcal{D}_x^2\right)(F\cdot G)&=&0,\label{HirotaDDEq1}\\
\frac12\mathcal{D}_x^2(G\cdot G)-\frac{\gamma}{N}\vert F\vert^2+\gamma\, G^2&=&0\label{HirotaDDEq2}.
 \end{eqnarray}
To obtain vector $M-$soliton solutions, we assume that the $\epsilon-$expansions (\ref{vectorexpan}) truncate to
\begin{equation}
    F=F_0+\sum_{k=1}^M\epsilon^{k}F_{k}\quad \mbox{and}\quad G=1+\sum_{k=1}^M\epsilon^{k}G_{k},\label{expanDD}
\end{equation}
where $F_0$ is the vector given in Eq.(\ref{BackDD}).

\subsubsection{The vector one-soliton solution}

For the vector one-soliton solution, we take $M=1$ in the $\epsilon-$expansions (\ref{expanDD}). Introducing these expansions in the bilinear equations (\ref{HirotaDDEq1}) and (\ref{HirotaDDEq2}), we get a new set of bilinear equations to solve for the components $F_1$ and $G_1$. To solve the equations associated to $\epsilon^1$, we make the ansatz:
\begin{equation}
    G_1=e^{\Lambda}\quad \mbox{and}\quad F_1=v_1\, e^{\Lambda},
\end{equation}
where $v_1\in \mathbb{C}^N$ is a constant vector and $\Lambda=\mu\, x+\lambda\, t+\varphi$. Here, $\mu$, $\lambda$ and $\varphi$ are real constants to determine. We get
\begin{equation}
    v_1=e^{i\theta}F_0,\quad \lambda=-\kappa\mu-\gamma\sin(\theta)\quad \mbox{and}\quad \mu^2=-4\gamma\sin^2\left(\frac{\theta}{2}\right),
\end{equation}
where $\theta\in\mathbb{R}$. The last constraint implies $\gamma<0$ and leads to repulsive interactions typical of dark soliton solutions for the nonlinear Schrödinger system (\ref{Master}). The equations associated with $\epsilon^2$ are solved using the ansatz and constraints mentioned above.

For $\epsilon=1$, we get the exact soliton solution
\begin{equation}
    \vert\Psi\vert^2=1-\sin^2\left(\frac{\theta}{2}\right)\mbox{sech}^2\left(\frac{\Lambda}{2}\right).\label{OneDarkVec}
\end{equation}

The vector one-soliton solution obtained in this regime corresponds to a dark vector soliton propagating on the finite-density background $\vert\Psi\vert^2=1$. Unlike the bright one-soliton solution of Sec.~IV.A, the density does not vanish asymptotically, but approaches a constant value fixed by the ground-state configuration. The localized density depression therefore represents a coherent excitation of the background rather than an isolated localized wave packet.

\begin{figure*}[h]
    \centering
    \includegraphics[width=0.7\textwidth]{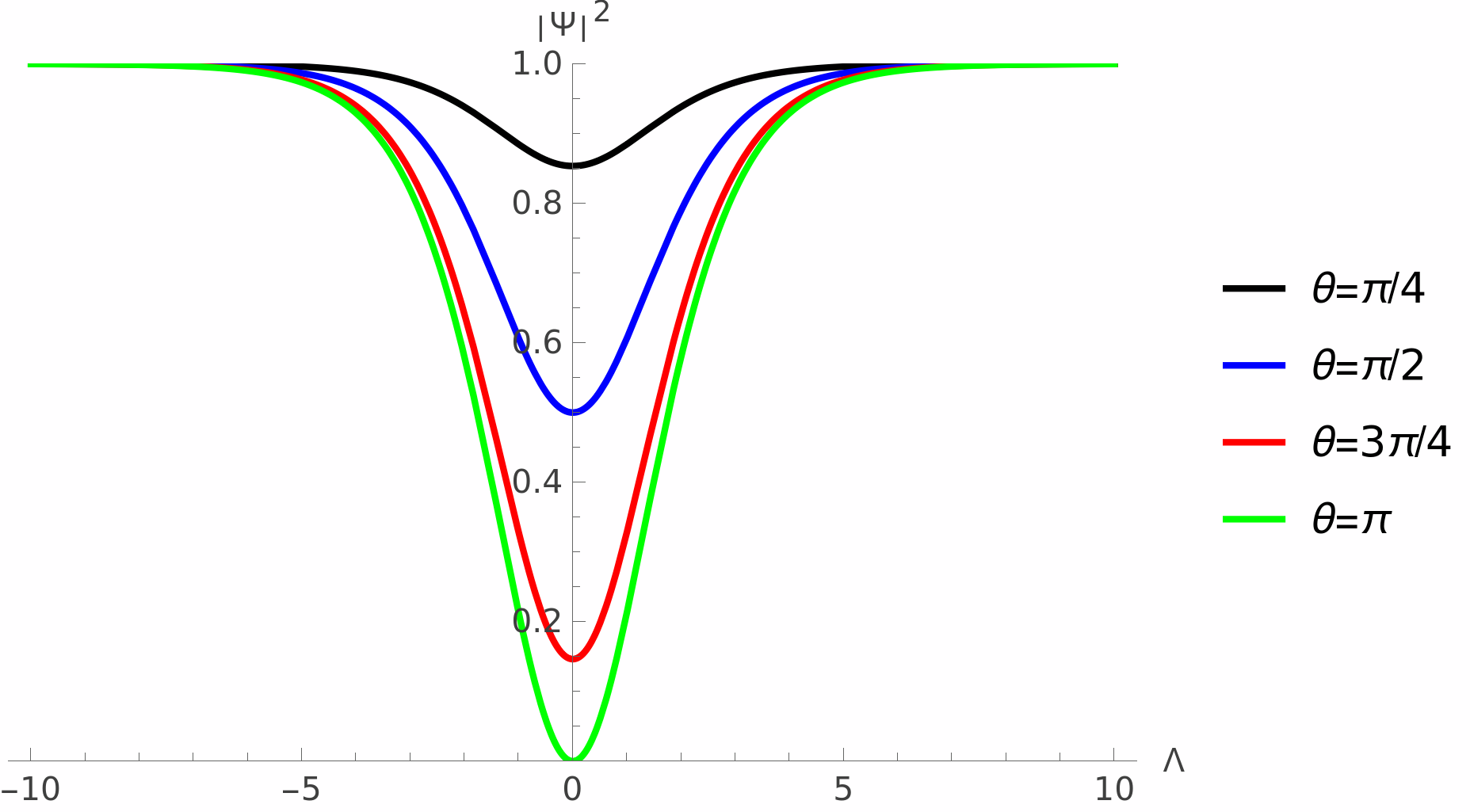}
    \caption{The density function $\vert\Psi(x,t)\vert^2$ given in Eq.(\ref{OneDarkVec}) as a function of $\Lambda$ for $\theta=\pi/4,\pi/2,3\pi/4,\pi$.}
    \label{VecDarkDepth}
\end{figure*}

In Figure~\ref{VecDarkDepth}, we show the effect of $\theta$ on the density depression of the vector dark soliton solution (\ref{OneDarkVec}). The persistence of a non-zero asymptotic density on both sides of the localized dip clearly distinguishes the present solution from the bright-soliton case. The density minimum propagates without deformation, illustrating the balance between dispersion and nonlinearity that characterizes integrable dark-soliton dynamics.

\subsubsection{The vector two-soliton solution}

For the vector two-soliton solution, we take $M=2$ in Eq.(\ref{expanDD}). For the equations associated to $\epsilon^1$, we assume that
\begin{equation}
    G_1=e^{\Lambda_1}+e^{\Lambda_2}\quad \mbox{and}\quad F_1=v_1\,e^{\Lambda_1}+v_2\,e^{\Lambda_2},
\end{equation}
where $v_1, v_2\in \mathbb{C}^N$ are constant vectors and $\Lambda_j=\mu_j\,x+\lambda_j\,t+\varphi_j$. We find
\begin{equation}
    v_j=e^{i\theta_j}F_0,\quad \lambda_j=-\kappa \mu_j-\gamma\sin(\theta_j)\quad \mbox{and}\quad \mu_j^2=-4\gamma\sin^2\left(\frac{\theta_j}{2}\right),\label{dispDD}
\end{equation}
where $\theta_1,\theta_2\in\mathbb{R}.$ For two distinct solitary waves, we assume $\theta_1,\theta_2\in(0,2\pi)$ such that $\theta_1\neq\theta_2$. For $F_2$ and $G_2$, we take
\begin{equation}
    G_2=\mathcal{A}_{12}\, e^{\Lambda_1+\Lambda_2}\quad \mbox{and}\quad F_2=v_{12}\, e^{\Lambda_1+\Lambda_2},
\end{equation}
where $\mathcal{A}_{12}\in\mathbb{R}$ and $v_{12}\in\mathbb{C}^N$ is a constant vector. We find
\begin{equation}
    \mathcal{A}_{12}=\frac{\sin^2\left(\frac{\theta_1-\theta_2}{4}\right)}{\sin^2\left(\frac{\theta_1+\theta_2}{4}\right)}\quad \mbox{and}                     \quad v_{12}=\mathcal{A}_{12}\,e^{i(\theta_1+\theta_2)}\,F_0.
\end{equation}
With the above expressions, the Hirota bilinear equations (\ref{HirotaDDEq1}) and (\ref{HirotaDDEq2}) are solved. We thus get an exact expression for the vector two-soliton solution. The coefficient $\mathcal{A}_{12}$ is used to quantify the phase-shift transition between solitons. The present authors in \cite{DelisleDark} made a thorough analysis of this coefficient for the Eq.(\ref{Master}) in the scalar case $N=1$. This coefficient remains in the coupled system of nonlinear Schrödinger equations.

\begin{figure*}[h]
    \centering
    \includegraphics[width=0.9\textwidth]{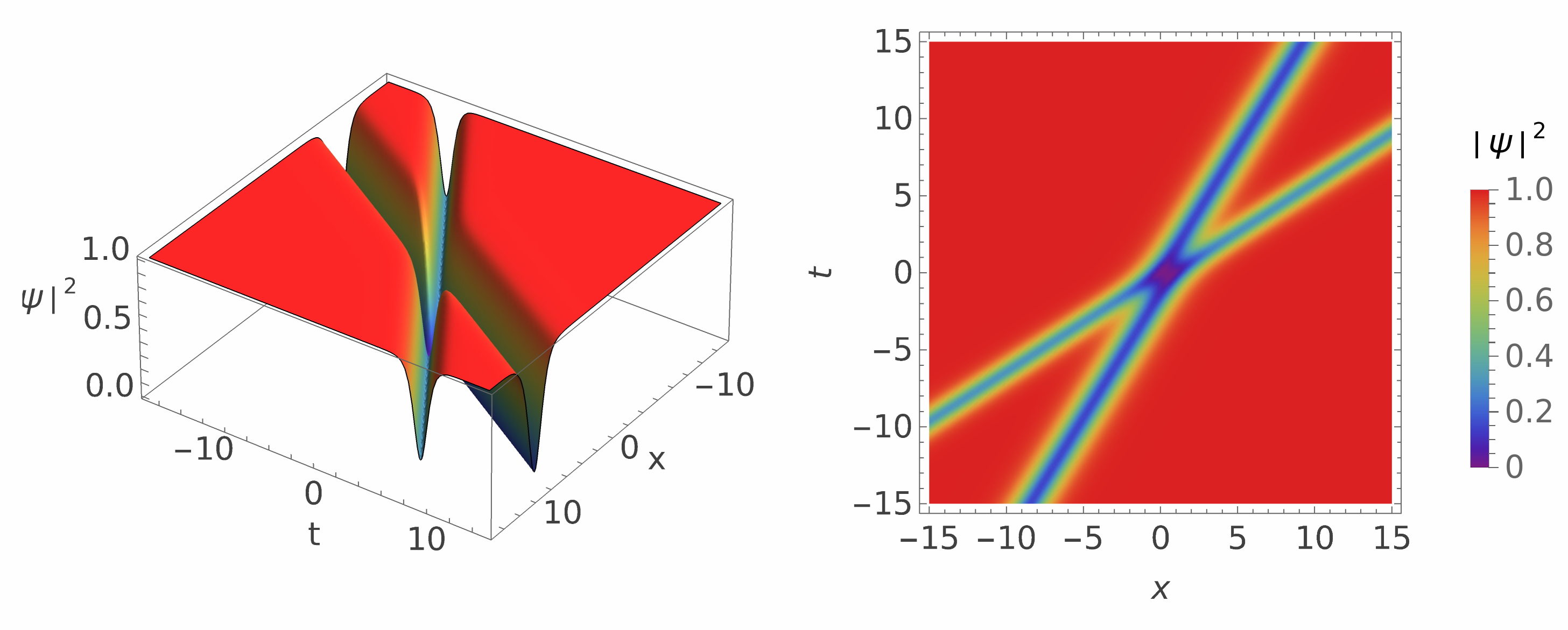}
    \caption{A vector $2-$soliton solution given by the density function $\vert\Psi(x,t)\vert^2$ in Eq.(\ref{DensityVec}) for $\gamma=-1$, $\kappa=\epsilon=1$, $\theta_1=3\pi/4$, $\theta_2=11\pi/8$, and $\varphi_1=\varphi_2=0$.}
    \label{DD2SoluionVec}
\end{figure*}

In Figure~\ref{DD2SoluionVec}, we present a vector two-soliton solution described by the density function $\vert\Psi\vert^2$ given in Eq.(\ref{DensityVec}) for $\gamma=-1$, $\kappa=\epsilon=1$, $\theta_1=3\pi/4$, $\theta_2=11\pi/8$, and  $\varphi_1=\varphi_2=0$. We can observe two solitary waves of different intensity depressions and velocity. The nonlinear interaction manifests itself as a phase shift transition at time $t\approx 0$.

As in the bright-soliton case, the two-soliton solution contains interaction-generated terms arising from the nonlinear coupling of two elementary vector excitations. However, because the solitons evolve on a finite-density background, the collision dynamics differs significantly from those of bright solitons. In particular, the most prominent nonlinear signature is no longer the appearance of localized interference fringes but rather the emergence of phase shifts and trajectory displacements after the interaction.

\subsubsection{The vector three-soliton}

The construction of the vector three-soliton solution provides a nontrivial validation of the vector Hirota formalism in the finite-background regime. As in the bright sector, the existence of an unconstrained three-soliton solution confirms that the vector bilinear representation preserves the complete integrability of the original multi-component nonlinear Schrödinger system.

For the vector three-soliton solution, we take $M=3$ in Eq.(\ref{expanDD}). Based on the traveling wave and the two-soliton vector solutions, we take the following forms for the components $F_j$ and $G_j$ of the functional expansions:
\begin{eqnarray}
    F_1&=&F_0 \,\sum_{k=1}^3e^{i\theta_k+\Lambda_k},\quad G_1=\sum_{k=1}^3e^{\Lambda_k},\\
    F_2&=&F_0\,\sum_{1\leq p<q\leq 3}\mathcal{A}_{pq}e^{i(\theta_p+\theta_q)+\Lambda_p+\Lambda_q},\quad G_2=\sum_{1\leq p<q\leq3}\mathcal{A}_{pq}e^{\Lambda_p+\Lambda_q},\\
    F_3&=&F_0\mathcal{A}_{123}e^{i(\theta_1+\theta_2+\theta_3)+\Lambda_1+\Lambda_2+\Lambda_3},\quad G_3=\mathcal{A}_{123}e^{\Lambda_1+\Lambda_2+\Lambda_3},
\end{eqnarray}
where $\Lambda_j=\mu_j\,x+\lambda_j\,t+\varphi_j$ and $\theta_j\in (0,2\pi)$ are real constants such that $\theta_1\neq\theta_2\neq\theta_3$. In order for $F$ and $G$ to be a solution of the bilinear equations (\ref{HirotaDDEq1}) and (\ref{HirotaDDEq2}), the constants $\lambda_j$ and $\mu_j$ must have the form (\ref{dispDD}), 
\begin{equation}
    \mathcal{A}_{pq}=\frac{\sin^2\left(\frac{\theta_p-\theta_q}{4}\right)}{\sin^2\left(\frac{\theta_p+\theta_q}{4}\right)}\quad \mbox{and}\quad \mathcal{A}_{123}=\mathcal{A}_{12}\mathcal{A}_{13}\mathcal{A}_{23}.
\end{equation}

\begin{figure*}[h]
    \centering
    \includegraphics[width=0.9\textwidth]{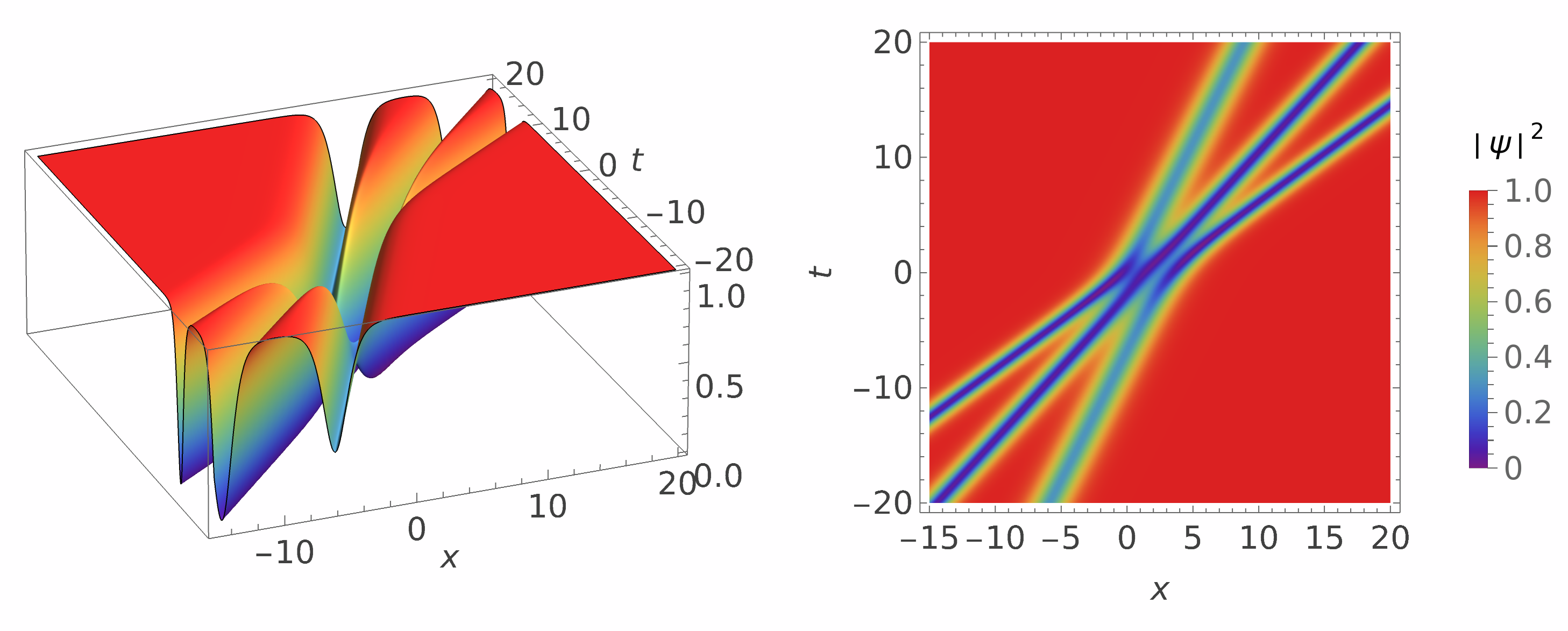}
    \caption{A vector $3-$soliton solution given by the density function $\vert\Psi(x,t)\vert^2$ in Eq.(\ref{DensityVec}) for $\gamma=-1$, $\kappa=\epsilon=1$, $\theta_1=5\pi/8$, $\theta_2=7\pi/8$, $\theta_3=9\pi/8$ and  $\varphi_1=\varphi_2=\varphi_3=0$.}
    \label{DD3SoluionVec}
\end{figure*}

In Figure~\ref{DD3SoluionVec}, we present a vector three-soliton solution described by the density function $\vert\Psi\vert^2$ given in Eq.(\ref{DensityVec}) for $\gamma=-1$, $\kappa=\epsilon=1$, $\theta_1=5\pi/8$, $\theta_2=7\pi/8$, $\theta_3=9\pi/8$ and  $\varphi_1=\varphi_2=\varphi_3=0$. 
The three localized density depressions propagate on a common finite-density background and undergo a sequence of pairwise interactions. In contrast to the bright three-soliton solution, the dominant nonlinear signature is not the formation of strong interference maxima but rather the cumulative phase shifts acquired during successive collisions. Despite these repeated interactions, each dark soliton preserves its identity and recovers its original profile after the collision process, demonstrating the robustness of multi-soliton dynamics in completely integrable systems. The dynamics shows strong phase shift manifestations of dark soliton collisions. This nonlinear signature of dark solitons interactions can be explicitly quantified \cite{DelisleDark} using the coefficients $A_{pq}$.


The dark one-, two-, and three-soliton solutions highlight the distinctive role played by the finite-density background in multi-component nonlinear Schrödinger systems. Whereas bright-soliton interactions are characterized by localized interference structures and transient density enhancement, dark-soliton interactions primarily manifest through phase shifts and trajectory displacements. The vector Hirota formalism successfully captures both behaviors within a unified framework while preserving the intrinsic multi-component structure of the model.

\subsection{Mixed vector soliton solutions}

In the case $\vert F_0\vert^2=N$, we have assumed that the vector $F_0$ was uniformly distributed among its components. The vector $F_0$ was chosen as in Eq.(\ref{BackDD}). However, the nonlinear constraint $\vert F_0\vert^2=N$ may be verified using a non-uniform background, such as
\begin{equation}
    F_0=\sqrt{\frac{N}{K}}\,\sum_{k=1}^Ke_k,
\end{equation}
which should generate mixed-type vector soliton: $(K)$ dark solitons together with $(N-K)$ bright solitons. Such non-uniform population distributions are closely related to phase-separated multi-component condensates \cite{Timmermans}. The Hirota bilinear equations to solved remain the same and are given by equations (\ref{HirotaDDEq1}) and (\ref{HirotaDDEq2}). We suppose $K\neq N$, this case has been studied in the previous section.

The mixed-type solitons constitute the most general class of solutions considered in this work. Unlike the bright and dark cases, which correspond respectively to uniform vanishing and finite-density backgrounds, the present construction allows only a subset of the vector components to carry a background density. As a consequence, bright and dark excitations coexist within the same vector state, leading to genuinely multi-component nonlinear structures
\cite{Busch,Ohberg2001,Kanna,Morera2018,Perez2005}. This coexistence highlights the flexibility of the vector Hirota formalism and provides a unified framework connecting the bright and dark soliton sectors.

\subsubsection{The mixed vector one soliton solution}

We assume the $\epsilon$-expansions:
\begin{equation}
    F=F_0+\epsilon F_1+\epsilon^2F_2\quad \mbox{and}\quad G=1+\epsilon^2 G_2.
\end{equation}
Using the same notation as in Section IV A., we take 
\begin{equation}
    F_1=v_1\, e^{\Xi},\quad F_2=v_2\, e^{\Xi+\Xi^*}\quad \mbox{and}\quad G_2=\alpha\, e^{\Xi+\Xi^*},
\end{equation}
where $v_1,v_2\in\mathbb{C}^N$ are constant complex vectors and $\alpha\in\mathbb{R}$ is a real constant. Introducing these finite expansions in the bilinear equations (\ref{HirotaDDEq1}) and (\ref{HirotaDDEq2}), we explicitly find:
\begin{equation}
    \Omega=-\kappa\Theta+\frac{i}{2}\Theta^2,\quad F_0^{\dagger}v_1=0\quad\mbox{and}\quad v_2=-\alpha\frac{\Theta^2}{\vert\Theta\vert^2}F_0,
\end{equation}
where
\begin{equation}
    \alpha=\frac{\gamma\vert\Theta\vert^2\vert v_1\vert^2}{N(\vert\Theta\vert^2+\gamma)(\Theta+\Theta^*)^2}.
\end{equation}

\begin{figure*}[h]
    \centering
    \includegraphics[width=0.6\textwidth,height=8cm]{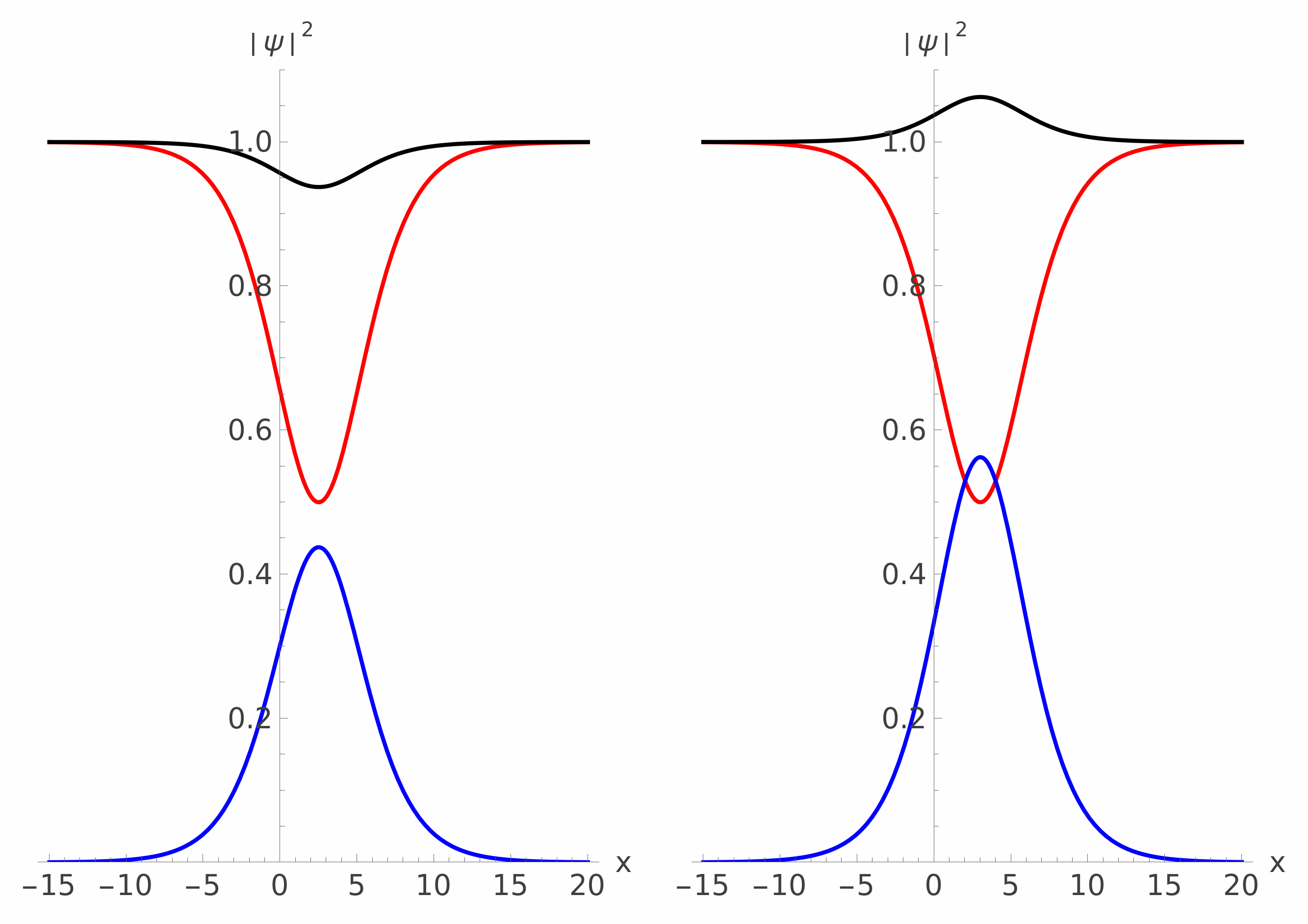}
    \caption{A mixed vector dark-bright and bright-dark one-soliton solution at $t=0$ given by the density function $\vert\Psi\vert^2=\vert\psi_1\vert^2+\vert\psi_2\vert^2$ in Eq.(\ref{DensityVec}) with $\kappa=\epsilon=1$, $F_0^t=(\sqrt{2},0)$, $v_1^t=(0,1)$, $\Phi=0$ and $\Theta=0.25-0.25i$. The black curve corresponds to $\vert\Psi\vert^2$, $\vert\psi_1\vert^2$ to the red one, while the blue curve is the density $\vert\psi_2\vert^2$. The left panel is for $\gamma=-1$ and the right panel is for $\gamma=1$.  }
    \label{ProofDarkBright}
\end{figure*}

We can show that $\vert\Psi\vert^2>1$ if and only if $\vert\Theta\vert^2>-\gamma$. This result can be observed in Figure~\ref{ProofDarkBright}. Indeed, the black curve is either above or below the line $\vert\Psi\vert^2=1$ which is the uniform density background. So, we deduce that for $\gamma>0$, the resulting vector soliton manifests as a lump. For $\gamma<0$, we may observe two distinct dynamics depending of $\vert\Theta\vert^2$.

The mixed one-soliton solution exhibits a coexistence of bright and dark characteristics within different components of the vector field\cite{Busch,Ohberg2001}. The orthogonality condition $F_0^\dagger v_1=0$ ensures that the localized excitation develops in a subspace complementary to the background configuration. Consequently, some components support a localized density enhancement while others exhibit a localized density depletion, resulting in a hybrid vector solitary wave.

To produce Figure~\ref{ProofDarkBright}, we take $N=2$ in the vector solution $\Psi$ given in Eq.(\ref{vectorPsi}). The red, blue and black curves correspond, respectively, to the density functions $\vert \psi_1\vert^2$, $\vert\psi_2\vert^2$ and $\vert \Psi\vert^2=\vert\psi_1\vert^2+\vert\psi_2\vert^2.$ In order to observe the interplay between the dark and bright soliton, we have taken $t=0$ as typical time. For both figures are for $\kappa=\epsilon=1$, $F_0^t=(\sqrt{2},0)$, $v_1^t=(0,1)$, $\Phi=0$ and $\Theta=0.25-0.25i$. The left figure is for $\gamma=-1$ and and the right one for $\gamma=1$.

For the left figure, the intensity depression of the dark soliton $\vert\psi_1\vert^2$ (red curve) is greater than the intensity lump of the bright soliton $\vert\psi_2\vert^2$ (blue curve). The resulting total density $\vert\Psi\vert^2$ as a dark dynamic manifestation which shows a dark-bright vector soliton.

For right panel in Figure~\ref{ProofDarkBright}, the total density $\vert\Psi\vert^2$ corresponds to a vector one-soliton solution of bright-dark type. The amplitude of the bright soliton (blue curve) is greater than the intensity depression of the dark soliton (red curve), resulting in a seemingly bright dynamic for the total density. 
Figure~\ref{ProofDarkBright} clearly illustrates the multi-component nature of the mixed vector soliton. Although individual components display qualitatively different behaviors, the total density results from their nonlinear superposition. Depending on the relative amplitudes of the bright and dark contributions, the resulting vector excitation may exhibit either a globally bright or a globally dark character.

\subsubsection{The mixed vector two soliton solution}

For a vector two-soliton solution, we assume the following $\epsilon$-expansions
\begin{equation}
    F=F_0+\epsilon F_1+\epsilon^2F_2+\epsilon^3 F_3+\epsilon^4 F_4\quad \mbox{and}\quad G=1+\epsilon^2G_2+\epsilon^4 G_4
\end{equation}
and take
\begin{equation}
    F_1=v_1\, e^{\Xi_1}+v_2\, e^{\Xi_2},
\end{equation}
where $v_1, v_2\in\mathbb{C}^N$. We get from the $\epsilon^1$ equation:
\begin{equation}
    \Omega_j=-\kappa\Theta_j+\frac{i}{2}\Theta_j^2\quad \mbox{and}\quad F_0^{\dagger}v_j=0
\end{equation}
for $j=1,2$. For the $\epsilon^2$ equations, we suppose the following forms for the components $F_2$ and $G_2$:
\begin{eqnarray}
   F_2&=&v_1^1\, e^{\Xi_1+\Xi_1^*}+v_1^2\, e^{\Xi_1+\Xi_2^*}+v_2^1\, e^{\Xi_2+\Xi_1^*}+v_2^2\, e^{\Xi_2+\Xi_2^*},\\
   G_2&=&\alpha_1^1\, e^{\Xi_1+\Xi_1^*}+\alpha_1^2\, e^{\Xi_1+\Xi_2^*}+\alpha_2^1\, e^{\Xi_2+\Xi_1^*}+\alpha_2^2\, e^{\Xi_2+\Xi_2^*},
\end{eqnarray}
where $v_p^q$ and $\alpha_p^q$ are, respectively, vectors of $\mathbb{C}^N$ and complex constants. They satisfy $(\alpha_p^q)^{\dagger}=\alpha_q^p$. We find
\begin{equation}
    v_p^q=-\alpha_p^q\frac{\Theta_p\Theta_q}{\vert\Theta_q\vert^2}F_0\quad \mbox{and}\quad\alpha_p^q=\frac{\gamma\,\Theta_q^{*}\Theta_p\, v_q^{\dagger}v_p}{N(\Theta_q^*\Theta_p+\gamma)(\Theta_p+\Theta_q^*)^2}.
\end{equation}
For the $\epsilon^3$ equation, we assume
\begin{equation}
    F_3=v_{12}^1\, e^{\Xi_1+\Xi_2+\Xi_1^*}+v_{12}^2\, e^{\Xi_1+\Xi_2+\Xi_2^*}
\end{equation}
and find
\begin{equation}
    v_{12}^p=A_{12}\left(\frac{\alpha_2^p}{\Theta_1+\Theta_p^*}v_1-\frac{\alpha_1^p}{\Theta_2+\Theta_p^*}v_2\right),\quad A_{12}=\Theta_1-\Theta_2,
\end{equation}
for $p=1,2$. For the $\epsilon^4$ equation, let us assume
\begin{equation}
    F_4=v_{12}^{12}\, e^{\Xi_1+\Xi_2+\Xi_1^*+\Xi_2^*}\quad \mbox{and}\quad G_4=\alpha_{12}^{12}\,e^{\Xi_1+\Xi_2+\Xi_1^*+\Xi_2^*}.
\end{equation}
We get
\begin{eqnarray}
    v_{12}^{12}&=&\alpha_{12}^{12}\frac{\Theta_1^2\Theta_2^2}{\vert\Theta_1\vert^2\vert\Theta_2\vert^2}F_0\quad\mbox{and}\\\alpha_{12}^{12}&=&\vert\Theta_1-\Theta_2\vert^2\left(\frac{\alpha_1^1\alpha_2^2}{(\Theta_1+\Theta_2^*)(\Theta_2+\Theta_1^*)}-\frac{\alpha_1^2\alpha_2^1}{(\Theta_1+\Theta_1^*)(\Theta_2+\Theta_2^*)}\right).
\end{eqnarray}
The above expressions are sufficient in order for the bilinear equations (\ref{HirotaDDEq1}) and (\ref{HirotaDDEq2}) to be solved.

The mixed vector two-soliton solution combines the interaction mechanisms observed separately in the bright and dark sectors. Because each vector soliton contains both bright and dark components, the resulting collision dynamics involves a simultaneous interplay of localized density enhancement and density depletion. The interaction therefore exhibits a richer structure than in the purely bright or purely dark cases.

\begin{figure*}[h]
    \centering
    \includegraphics[width=0.9\textwidth]{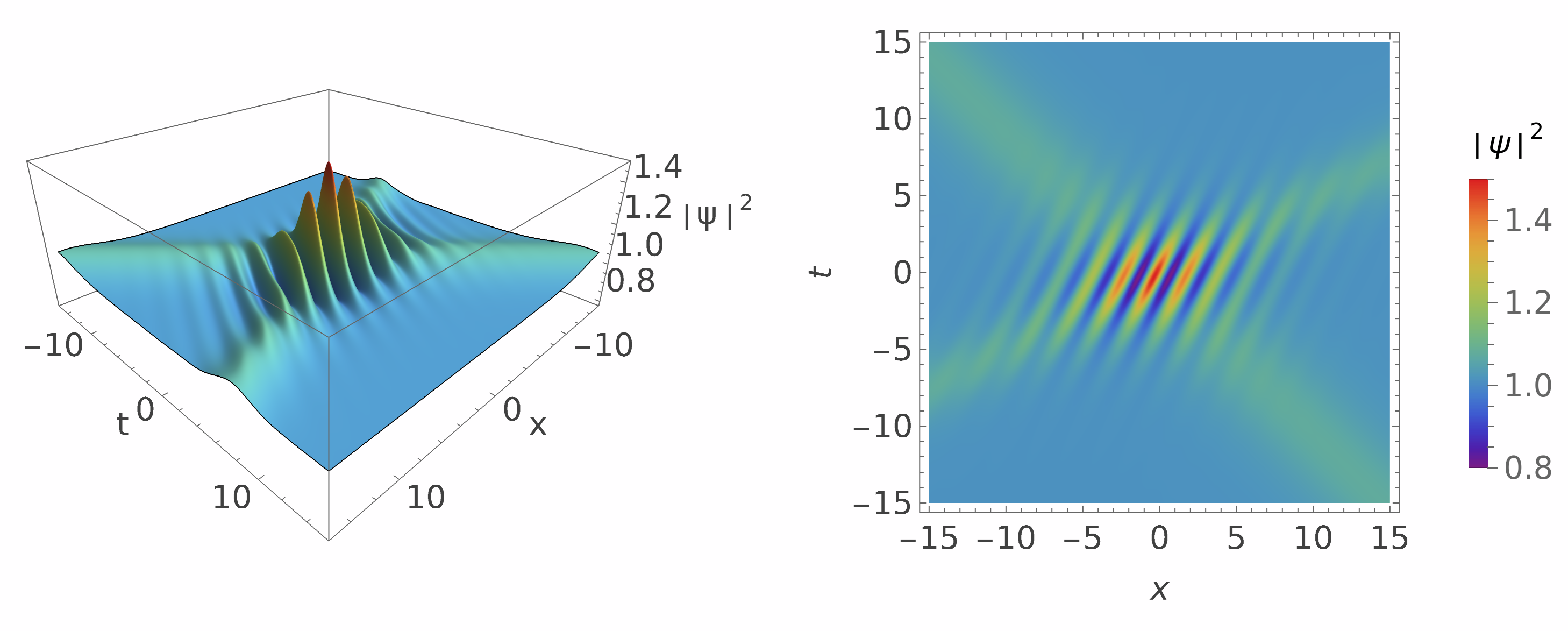}
    \caption{A mixed vector two-soliton solution given by the density function $\vert\Psi(x,t)\vert^2$ in Eq.(\ref{DensityVec}) for $\gamma=1$, $\kappa=\epsilon=1$, $F_0^t=(\sqrt{2},0)$, $v_1^t=v_2^t=(0,1)$ and $\Phi_1=\Phi_2=0$. The spectral parameters are  $(\Theta_1,\Theta_2)=(0.25-2i;0.25+i)$. }
   \label{Mixed2vector}
\end{figure*}

\begin{figure*}[h]
    \centering
    \includegraphics[width=0.9\textwidth]{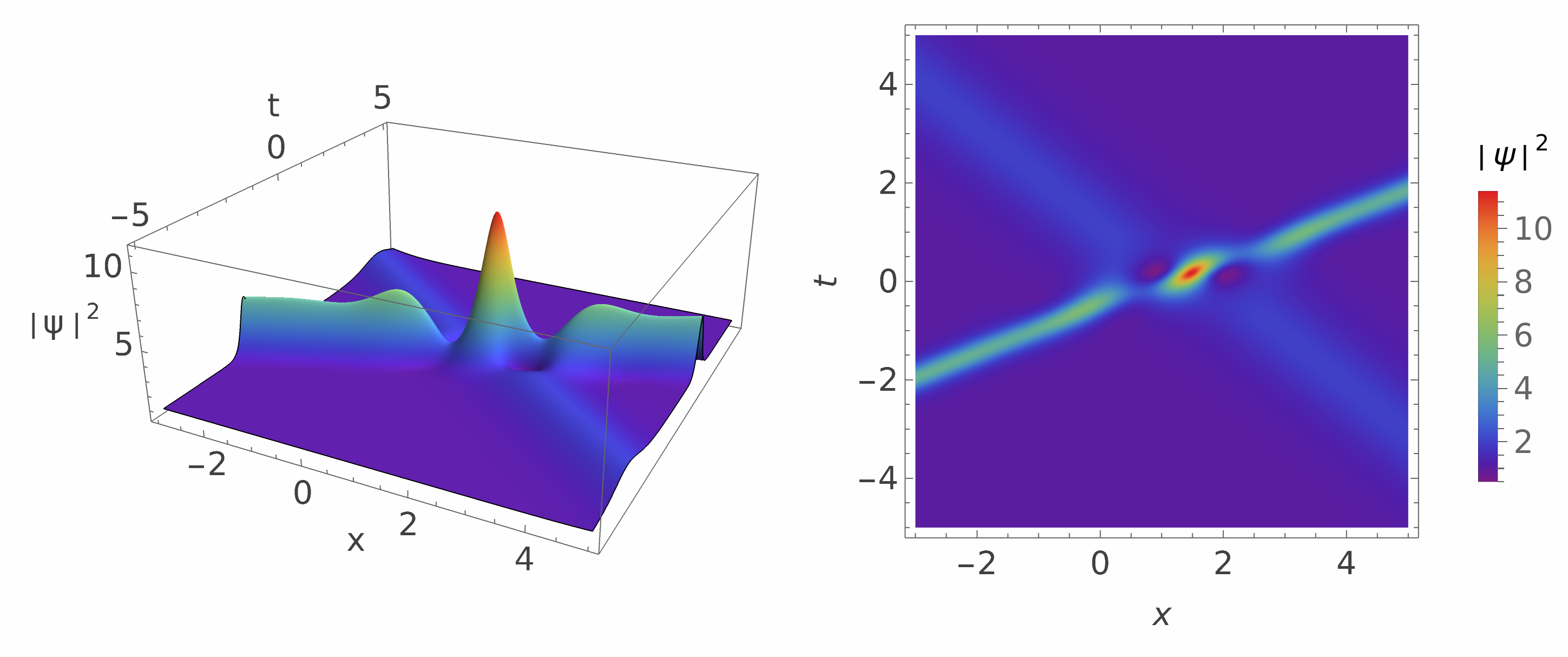}
    \caption{A mixed vector two-soliton solution given by the density function $\vert\Psi(x,t)\vert^2$ in Eq.(\ref{DensityVec}) for $\gamma=1$, $\kappa=\epsilon=1$, $F_0^t=(\sqrt{2},0)$, $v_1^t=v_2^t=(0,1)$ and $\Phi_1=\Phi_2=0$. The spectral parameters are  $(\Theta_1,\Theta_2)=(1-2i;2+i)$. }
   \label{Mixed2vector1}
\end{figure*}

In Figure~\ref{Mixed2vector} and Figure~\ref{Mixed2vector1}, we show two distinct vector two-soliton solutions of mixed type. In Figure~\ref{Mixed2vector}, we can identify strong interference fringes. We can see the dark-bright interplay as the density function goes under and above the uniform background $\vert \Psi\vert^2=\vert F_0\vert^2/N^2=1$. In Figure~\ref{Mixed2vector1}, we can observe a different dynamic for the dark-bright two-soliton solution. The two apparent lumps plunge in a deep before interaction creating a strong density pic. They retrieve their identities as one expects. We can also observe the dark contribution as the vector soliton manifest phase-shift transition.



\subsubsection{The mixed vector three-soliton solution}
The mixed three-soliton solution provides the most general nonlinear excitation considered in the present work. It combines the complexity of multi-soliton interactions with the coexistence of bright and dark components within the same vector state. The existence of an exact three-soliton solution in this regime further confirms that the vector Hirota formulation preserves integrability across all classes of vector soliton configurations.

To generalize the one and two mixed type soliton solution, we assume that the complex vector $F\in \mathcal{V}_D\oplus\mathcal{V}_B$ where $\mathcal{V}_D$ and $\mathcal{V}_B$ are vector subspaces. We assume that $\mathcal{V}_D=\mbox{span}(F_0)$ is the subspace in which the dark soliton part will belong, whereas $\mathcal{V}_B\subseteq\mathcal{V}_D^{\perp}$ is the bright subspace that we assume orthogonal to the dark soliton subspace. We write
\begin{equation}
    F=F_0\, f_D+F_B,
\end{equation}
where $f_D$ is a complex scalar function that describes the dark part of the vector soliton solution and $F_B$ is a complex vector function that represents the bright contribution. By construction, we have $F_B^{\dagger}F_0=0$. Introducing this orthogonal decomposition in the Hirota bilinear equations (\ref{HirotaDDEq1}) and (\ref{HirotaDDEq2}) yields
\begin{eqnarray}
    \left(i\mathcal{D}_t+i\kappa\mathcal{D}_x+\frac12\mathcal{D}_x^2\right)(F_B\cdot G)&=&0,\\
    \left(i\mathcal{D}_t+i\kappa\mathcal{D}_x+\frac12\mathcal{D}_x^2\right)(f_D\cdot G)&=&0\\
    \frac12\mathcal{D}_x^2(G\cdot G)-\gamma \vert f_D\vert^2-\frac{\gamma}{N}\vert F_B\vert^2+\gamma G^2&=&0.
\end{eqnarray}
To obtain a vector three-soliton solution, we assume the $\epsilon-$expansions
\begin{equation}
    f_D=1+\epsilon^2f_2+\epsilon^4f_4+\epsilon^6f_6,\quad
    F_B=\epsilon F_1+\epsilon^3 F_3+\epsilon^5F_5,\quad
    G=1+\epsilon^2G_2+\epsilon^4G_4+\epsilon^6G_6,\label{3SolComMix}
\end{equation}
where the functions $f_{2k}$ and $G_{2k}$ are, respectively, complex and real-valued scalar functions, and $F_{2k-1}$ are complex-valued vector functions for $k=1,2,3.$

 For a vector three-soliton solution, we assume 
\begin{equation}
    F_1=v_1e^{\Xi_1}+v_2e^{\Xi_2}+v_3e^{\Xi_3},\label{3SolF1Mix}
\end{equation}
where $v_m\in \mathbb{C}^N$ are constant vectors. We get the constraints
\begin{equation}
    v_m^{\dagger}F_0=0\quad \mbox{and}\quad \Omega_m=-\kappa \Theta_m+\frac{i}{2}\Theta_m^2,
\end{equation}
for $m=1,2,3$. We thus assume that the bright subspace $\mathcal{V}_B=\mbox{span}(v_1,v_2,v_3)$. To keep the main text concise, the expressions of the other components are provided in the appendix~\ref{appendixA2}.

\begin{figure*}[h]
  \centering
    \includegraphics[width=0.85\textwidth]{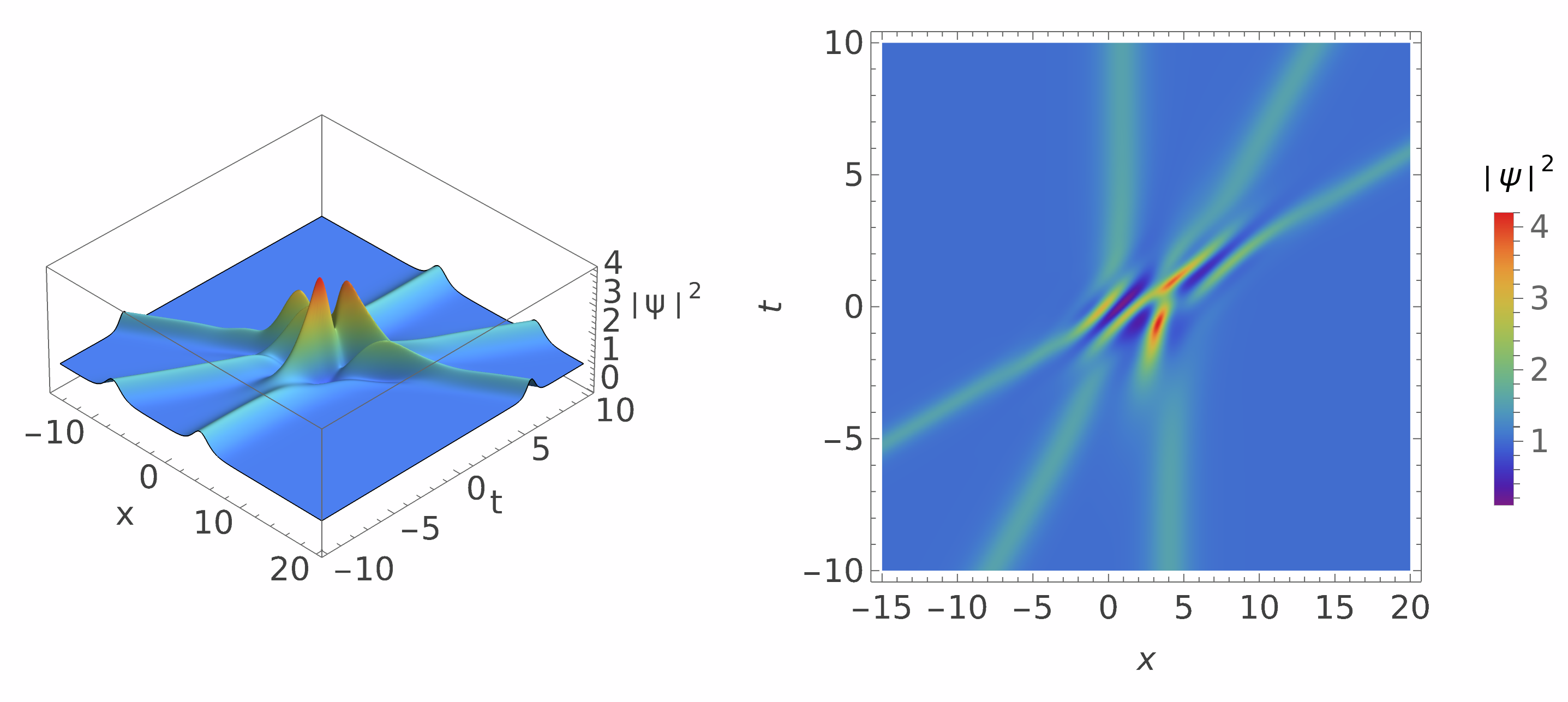}
    \caption{A mixed vector three-soliton solution given by the density function $\vert\Psi(x,t)\vert^2$ in Eq.(\ref{DensityVec}) for $\gamma=\kappa=\epsilon=1$, $F_0^t=(\sqrt{2},0)$, $v_1^t=v_2^t=v_3^t=(0,1)$ and $\Phi_1=\Phi_2=\Phi_3=0$. The spectral parameters are $\Theta_1=0.75+2i$, $\Theta_2=0.75-i$ and $\Theta_3=0.75$.}
    \label{3DarkBrightVector}
\end{figure*}

Figure~\ref{3DarkBrightVector} presents a mixed vector three-soliton solution. The three localized excitations propagate with distinct velocities and undergo successive nonlinear interactions while preserving their composite bright-dark character. The coexistence of density peaks and density depressions within the same vector structure generates a rich interaction pattern that combines features of both bright- and dark-soliton dynamics. Despite the complexity of these interactions, each soliton recovers its original profile after collision, illustrating the robustness of mixed vector soliton states and the complete integrability of the underlying system.

The mixed one-, two-, and three-soliton solutions provide a continuous bridge between the purely bright and purely dark sectors. They demonstrate that the vector Hirota formalism naturally accommodates hybrid nonlinear excitations in which localized density peaks and density depressions coexist within the same multi-component state. This unified description highlights the ability of the vector framework to capture a broad variety of coherent structures while preserving the symmetry and integrability properties of the original multi-component nonlinear Schrödinger system.

\section{Conclusion and perspectives}

In this work, we have developed a vector formulation of the Hirota bilinear formalism for the completely integrable multi-component nonlinear Schrödinger equation of Manakov type. Unlike the conventional component-wise approach, the proposed framework preserves the intrinsic vector structure and the underlying $U(N)$ symmetry of the model throughout the bilinearization procedure. This provides a compact representation of multi-component nonlinear excitations directly at the vector level. Using this formalism, we have constructed exact one-, two-, and three-soliton solutions in the bright, dark, and mixed sectors. Bright vector solitons were shown to exhibit characteristic interference structures and elastic collisions of integrable systems, whereas dark vector solitons display collision-induced phase shifts and trajectory displacements on a finite-density background. The mixed sector naturally bridges these two regimes and demonstrates the coexistence of bright and dark excitations within the same vector state. The existence of exact three-soliton solutions in all sectors further confirms that the vector bilinear representation preserves the complete integrability of the original model.
The present work highlights the advantages of treating multi-component integrable systems directly in vector form and provides a unified framework for studying coherent structures in coupled nonlinear media. It provides a natural bridge between mathematical integrability and physically relevant multi-component matter-wave excitations encountered in Bose--Einstein condensates and vector nonlinear optical systems
\cite{Trippenbach2000,Kevrekidis2016,Mollenauer,Hasegawa}. Future developments will focus on extending the vector Hirota formalism to other classes of exact solutions, including rogue waves, rational solutions, and beating stripe soliton solutions \cite{DingPRE}

\begin{acknowledgments}
This work did not receive external funding.
\end{acknowledgments}
\section*{AUTHOR DECLARATIONS}
\subsection*{Conflict of Interest}
The authors have no conflicts to disclose.
\subsection*{Authors Contributions}
\textbf{Laurent Delisle:} Conceptualization (equal); Formal analysis (equal); Investigation (equal); Methodology (equal); Writing – review \& editing (equal). \textbf{Amine Jaouadi:} Conceptualization (equal); Formal analysis (equal); Investigation (equal); Methodology (equal); Supervision (equal); Writing – review \& editing (equal).

\section*{Data Availability Statement}
Data sharing is not applicable to this article as no new data were created or analyzed in this study.





\appendix

\section{Vector three-soliton exact expressions}
In this appendix, we give the explicit expressions of the vector three-soliton solution in the bright and mixed case. The components of the functions $F$ and $G$ are given, as well as the different interaction coefficients involved in the construction.
\subsection{Vector bright three-soliton}\label{appendixA1}
For the bright type vector three-soliton solution, the components of the functions $F$ and $G$ in Eq.(\ref{expanBB}) for $M=3$ are given by Eq.(\ref{3SolF1BB}) and by
\begin{eqnarray}
    G_2&=&a_1^1e^{\Xi_1+\Xi_1^*}+a_1^2e^{\Xi_1+\Xi_2^*}+a_1^3e^{\Xi_1+\Xi_3^*}+a_2^1e^{\Xi_2+\Xi_1^*}+a_2^2e^{\Xi_2+\Xi_2^*}+a_2^3e^{\Xi_2+\Xi_3^*}+a_3^1e^{\Xi3+\Xi_1^*}\\
    &+&a_3^2e^{\Xi_3+\Xi_2^*}+a_3^3e^{\Xi_3+\Xi_3^*}\nonumber,\\
    F_3&=&v_{12}^1e^{\Xi_1+\Xi_2+\Xi_1^*}+v_{12}^2e^{\Xi_1+\Xi_2+\Xi_2^*}+v_{12}^3e^{\Xi_1+\Xi_2+\Xi_3^*}+v_{13}^1e^{\Xi_1+\Xi_3+\Xi_1^*}+v_{13}^2e^{\Xi_1+\Xi_3+\Xi_2^*}\\&+&v_{13}^3e^{\Xi_1+\Xi_3+\Xi_3^*}
    +v_{23}^1e^{\Xi_2+\Xi_3+\Xi_1^*}+v_{23}^2e^{\Xi_2+\Xi_3+\Xi_2^*}+v_{23}^3e^{\Xi_2+\Xi_3+\Xi_3^*},\nonumber\\
    G_4&=&a_{12}^{12}e^{\Xi_1+\Xi_2+\Xi_1^*+\Xi_2^*}+a_{12}^{13}e^{\Xi_1+\Xi_2+\Xi_1^*+\Xi_3^*}+a_{12}^{23}e^{\Xi_1+\Xi_2+\Xi_2^*+\Xi_3^*}+a_{13}^{12}e^{\Xi_1+\Xi_3+\Xi_1^*+\Xi_2^*}\\&+&a_{13}^{13}e^{\Xi_1+\Xi_3+\Xi_1^*+\Xi_3^*}+a_{13}^{23}e^{\Xi_1+\Xi_3+\Xi_2^*+\Xi_3^*}+a_{23}^{12}e^{\Xi_2+\Xi_3+\Xi_1^*+\Xi_2^*}+a_{23}^{13}e^{\Xi_2+\Xi_3+\Xi_1^*+\Xi_3^*}\nonumber\\&+&a_{23}^{23}e^{\Xi_2+\Xi_3+\Xi_2^*+\Xi_3^*}\nonumber\\
    F_5&=&v_{123}^{12}e^{\Xi_1+\Xi_2+\Xi_3+\Xi_1^*+\Xi_2^*}+v_{123}^{13}e^{\Xi_1+\Xi_2+\Xi_3+\Xi_1^*+\Xi_3^*}+v_{123}^{23}e^{\Xi_1+\Xi_2+\Xi_3+\Xi_2^*+\Xi_3^*},\\
    G_6&=&a_{123}^{123}e^{\Xi_1+\Xi_2+\Xi_3+\Xi_1^*+\Xi_2^*+\Xi_3^*},
\end{eqnarray}
where the constants $a_p^q$ are given in Eq.(\ref{apq}). For the other constants, we find
\begin{eqnarray}
    v_{pm}^n&=&A_{pm}\left(\frac{a_m^n}{\Theta_p+\Theta_n^*}v_p-\frac{a_p^n}{\Theta_m+\Theta_n^*}v_m\right),\quad A_{pm}=\Theta_p-\Theta_m,\\
    a_{mn}^{pq}&=&A_{mn}A^*_{pq}\left(\frac{a_m^pa_n^q}{(\Theta_n+\Theta_p^*)(\Theta_m+\Theta_q^*)}-\frac{a_m^qa_n^p}{(\Theta_m+\Theta_p^*)(\Theta_n+\Theta_q^*)}\right),\\
    v_{123}^{pq}&=&\frac{A_{12}A_{13}a_{23}^{pq}}{(\Theta_1+\Theta_p^*)(\Theta_1+\Theta_q^*)}v_1+\frac{A_{21}A_{23}a_{13}^{pq}}{(\Theta_2+\Theta_p^*)(\Theta_2+\Theta_q^*)}v_2+\frac{A_{31}A_{32}a_{12}^{pq}}{(\Theta_3+\Theta_p^*)(\Theta_3+\Theta_q^*)}v_3,\\
    a_{123}^{123}&=&\frac{\vert A_{13}\vert^2 A_{12}A_{23}^*a_1^3a_{23}^{12}}{(\Theta_1+\Theta_1^*)(\Theta_1+\Theta_2^*)(\Theta_2+\Theta_3^*)(\Theta_3+\Theta_3^*)}\\&-&\frac{\vert A_{12}\vert^2 A_{13}A_{23}^*a_1^2a_{23}^{13}}{(\Theta_1+\Theta_1^*)(\Theta_1+\Theta_3^*)(\Theta_2+\Theta_2^*)(\Theta_3+\Theta_2^*)}
    +\frac{\vert A_{13}\vert^2 \vert A_{12}\vert^2a_1^1a_{23}^{23}}{(\Theta_1+\Theta_2^*)(\Theta_1+\Theta_3^*)(\Theta_2+\Theta_1^*)(\Theta_3+\Theta_1^*)}.\nonumber
\end{eqnarray}

\subsection{Vector mixed three-soliton}
\label{appendixA2}
For the mixed type vector three-soliton solution, the components of the functions $F$ and $G$ in (\ref{3SolComMix}) are given by Eq.(\ref{3SolF1Mix}) and by
\begin{eqnarray}
    G_2&=&\alpha_1^1e^{\Xi_1+\Xi_1^*}+\alpha_1^2e^{\Xi_1+\Xi_2^*}+\alpha_1^3e^{\Xi_1+\Xi_3^*}+\alpha_2^1e^{\Xi_2+\Xi_1^*}+\alpha_2^2e^{\Xi_2+\Xi_2^*}+\alpha_2^3e^{\Xi_2+\Xi_3^*}\\&+&\alpha_3^1e^{\Xi_3+\Xi_1^*}+\alpha_3^2e^{\Xi_3+\Xi_2^*}+\alpha_3^3e^{\Xi_3+\Xi_3^*}\nonumber,\\
    f_2&=&\beta_1^1e^{\Xi_1+\Xi_1^*}+\beta_1^2e^{\Xi_1+\Xi_2^*}+\beta_1^3e^{\Xi_1+\Xi_3^*}+\beta_2^1e^{\Xi_2+\Xi_1^*}+\beta_2^2e^{\Xi_2+\Xi_2^*}+\beta_2^3e^{\Xi_2+\Xi_3^*}\\&+&\beta_3^1e^{\Xi_3+\Xi_1^*}+\beta_3^2e^{\Xi_3+\Xi_2^*}+\beta_3^3e^{\Xi_3+\Xi_3^*}\nonumber.\\
    F_3&=&v_{12}^1e^{\Xi_1+\Xi_2+\Xi_1^*}+v_{12}^2e^{\Xi_1+\Xi_2+\Xi_2^*}+v_{12}^3e^{\Xi_1+\Xi_2+\Xi_3^*}+v_{13}^1e^{\Xi_1+\Xi_3+\Xi_1^*}+v_{13}^2e^{\Xi_1+\Xi_3+\Xi_2^*}\\&+&v_{13}^3e^{\Xi_1+\Xi_3+\Xi_3^*}+v_{23}^1e^{\Xi_2+\Xi_3+\Xi_1^*}+v_{23}^2e^{\Xi_2+\Xi_3+\Xi_2^*}+v_{23}^3e^{\Xi_2+\Xi_3+\Xi_3^*}\nonumber\\
        G_4&=&\alpha_{12}^{12}e^{\Xi_1+\Xi_2+\Xi_1^*+\Xi_2^*}+\alpha_{12}^{13}e^{\Xi_1+\Xi_2+\Xi_1^*+\Xi_3^*}+\alpha_{12}^{23}e^{\Xi_1+\Xi_2+\Xi_2^*+\Xi_3^*}+\alpha_{13}^{12}e^{\Xi_1+\Xi_3+\Xi_1^*+\Xi_2^*}\\&+&\alpha_{13}^{13}e^{\Xi_1+\Xi_3+\Xi_1^*+\Xi_3^*}+\alpha_{13}^{23}e^{\Xi_1+\Xi_3+\Xi_2^*+\Xi_3^*}+\alpha_{23}^{12}e^{\Xi_2+\Xi_3+\Xi_1^*+\Xi_2^*}+\alpha_{23}^{13}e^{\Xi_2+\Xi_3+\Xi_1^*+\Xi_3^*}\nonumber\\&+&\alpha_{23}^{23}e^{\Xi_2+\Xi_3+\Xi_2^*+\Xi_3^*}\nonumber\\
    f_4&=&\beta_{12}^{12}e^{\Xi_1+\Xi_2+\Xi_1^*+\Xi_2^*}+\beta_{12}^{13}e^{\Xi_1+\Xi_2+\Xi_1^*+\Xi_3^*}+\beta_{12}^{23}e^{\Xi_1+\Xi_2+\Xi_2^*+\Xi_3^*}+\beta_{13}^{12}e^{\Xi_1+\Xi_3+\Xi_1^*+\Xi_2^*}\\&+&\beta_{13}^{13}e^{\Xi_1+\Xi_3+\Xi_1^*+\Xi_3^*}+\beta_{13}^{23}e^{\Xi_1+\Xi_3+\Xi_2^*+\Xi_3^*}+\beta_{23}^{12}e^{\Xi_2+\Xi_3+\Xi_1^*+\Xi_2^*}+\beta_{23}^{13}e^{\Xi_2+\Xi_3+\Xi_1^*+\Xi_3^*}\nonumber\\&+&\beta_{23}^{23}e^{\Xi_2+\Xi_3+\Xi_2^*+\Xi_3^*}\nonumber\\
    F_5&=&v_{123}^{12}e^{\Xi_1+\Xi_2+\Xi_3+\Xi_1^*+\Xi_2^*}+v_{123}^{13}e^{\Xi_1+\Xi_2+\Xi_3+\Xi_1^*+\Xi_3^*}+v_{123}^{23}e^{\Xi_1+\Xi_2+\Xi_3+\Xi_2^*+\Xi_3^*},\\
    G_6&=&\alpha_{123}^{123}e^{\Xi_1+\Xi_2+\Xi_3+\Xi_1^*+\Xi_2^*+\Xi_3^*},\\
    f_6&=&\beta_{123}^{123}e^{\Xi_1+\Xi_2+\Xi_3+\Xi_1^*+\Xi_2^*+\Xi_3^*},
\end{eqnarray}
where the constants are explicitly given as
\begin{eqnarray}
    \alpha_p^m&=&\frac{\gamma\Theta_m^*\Theta_p\, v_m^{\dagger}v_p}{N(\gamma+\Theta_m^*\Theta_p)(\Theta_m^*+\Theta_p)^2},\quad \beta_p^m=-\frac{\Theta_p}{\Theta_m^*}\alpha_p^m,\\
    v_{pq}^m&=&A_{pq}\left(\frac{\alpha_q^m}{\Theta_p+\Theta_m^*}v_p-\frac{\alpha_p^m}{\Theta_q+\Theta_m^*}v_q\right),\quad A_{pq}=\Theta_p-\Theta_q,\\
    \alpha_{mn}^{pq}&=&A_{mn}A_{pq}^*\left(\frac{\alpha_m^p\alpha_n^q}{(\Theta_n+\Theta_p^*)(\Theta_m+\Theta_q^*)}-\frac{\alpha_m^q\alpha_n^p}{(\Theta_m+\Theta_p^*)(\Theta_n+\Theta_q^*)}\right),\quad \beta_{mn}^{pq}=\frac{\Theta_m\Theta_n}{\Theta_p^*\Theta_q^*}\alpha_{mn}^{pq},\\
    v_{123}^{pq}&=&\frac{A_{12}A_{13}\alpha_{23}^{pq}}{(\Theta_1+\Theta_p^*)(\Theta_1+\Theta_q^*)}v_1-\frac{A_{12}A_{23}\alpha_{13}^{pq}}{(\Theta_2+\Theta_p^*)(\Theta_2+\Theta_q^*)}v_2+\frac{A_{13}A_{23}\alpha_{12}^{pq}}{(\Theta_3+\Theta_p^*)(\Theta_3+\Theta_q^*)}v_3,\\
    \alpha_{123}^{123}&=&\frac{\vert A_{13}\vert^2 A_{12}A_{23}^*\alpha_1^3\alpha_{23}^{12}}{(\Theta_1+\Theta_1^*)(\Theta_1+\Theta_2^*)(\Theta_2+\Theta_3^*)(\Theta_3+\Theta_3^*)}\\&-&\frac{\vert A_{12}\vert^2 A_{13}A_{23}^*\alpha_1^2\alpha_{23}^{13}}{(\Theta_1+\Theta_1^*)(\Theta_1+\Theta_3^*)(\Theta_2+\Theta_2^*)(\Theta_3+\Theta_2^*)}
    +\frac{\vert A_{13}\vert^2 \vert A_{12}\vert^2\alpha_1^1\alpha_{23}^{23}}{(\Theta_1+\Theta_2^*)(\Theta_1+\Theta_3^*)(\Theta_2+\Theta_1^*)(\Theta_3+\Theta_1^*)}\nonumber,\\
    \beta_{123}^{123}&=&-\frac{\Theta_1\Theta_2\Theta_3}{\Theta_1^*\Theta_2^*\Theta_3^*}\alpha_{123}^{123}.
\end{eqnarray}


\begin{thebibliography}{99}

\bibitem{Becker2008}
C.~Becker \emph{et al.},
``Oscillations and interactions of dark and dark--bright solitons in Bose--Einstein condensates,''
\emph{Nature Physics} \textbf{4}, 496--501 (2008).


\bibitem{DelisleDark}
L. Delisle and A. Jaouadi, Analytical control of multidark soliton collisions and  dynamics with tunable properties. \textit{AVS Quantum Sci.} \textbf{7}, 041402 (2025).

\bibitem{DelisleSym}
L. Delisle and A. Jaouadi, Symmetry-Driven Multi-Soliton Dynamics in Bose-Einstein Condensates in Reduced Dimensions. \textit{Symmetry} \textbf{17}(4), 582 (2025).

\bibitem{Celanie}
E. Célanie, L. Delisle and A. Jaouadi, Optically tuned soliton dynamics in Bose-Einstein condensates within dark traps. \textit{J. Phys. A: Math. Theor.} \textbf{57}, 485701 (2024).

\bibitem{Ivancevic}
V. G. Ivancevic, Adaptive-wave alternative for the Black-Scholes option pricing model. \textit{Cogn Comput} \textbf{2}, 17-30 (2010).

\bibitem{IvanDel}
L. Delisle, D. P. Hi and A. Jaouadi, Bright solitons in a regulatory Ivancevic model (RIM): analytical and numerical insights into price-wave dynamics. \textit{Nonlinear Dyn} \textbf{114}, 329 (2026).

\bibitem{Olver} P. J.~Olver, \emph{Applications of Lie Groups to Differential Equations} (Springer New York, NY, 1993)

\bibitem{Ablowitz} M. J.~Ablowtiz and H.~Segur, \emph{Solitons and the Inverse Scattering Transform} (SIAM, 1991)

\bibitem{Rogers} C.~Rogers and W. K.~Schief, \emph{Bäcklund and Darboux Transformations: Geometry and Modern Applications in Soliton Theory} (Cambridge University Press, 2002)

\bibitem{Hirota}
R.~Hirota,
\emph{The Direct Method in Soliton Theory} (Cambridge University Press, 2004).

\bibitem{Hietarinta}
J.~Hietarinta,
``Hirota's bilinear method and its connection with integrability,''
in \emph{Integrability}, Lecture Notes in Physics Vol.~767, ed.~A.~V.~Mikhailov (Springer, 2009).

\bibitem{Hietarinta1} J. Hietarinta, A search for bilinear equations passing Hirota's three soliton condition. I. KdV-type bilinear equations. \textit{J. Math. Phys.} \textbf{28}, 1732-1742 (1987).

\bibitem{Hietarinta2} J. Hietarinta: A search for bilinear equations passing Hirota's three soliton condition. II. mKdV-type bilinear equations. \textit{J. Math. Phys.} \textbf{28}, 2094-2101 (1987).

\bibitem{Hietarinta3}
J. Hietarinta: A search for bilinear equations passing Hirota's three-soliton condition. III. Sine-Gordon-type bilinear equations. \textit{J. Math. Phys.} \textbf{28}, 2586-2592 (1987).

\bibitem{Hietarinta4}
J. Hietarinta: A search for bilinear equations Hirota's three-soliton condition. IV. Complex bilinear equations. \textit{J. Math. Phys.} \textbf{29}, 628-635 (1988).

\bibitem{LiLi}
L. Li, C. Duan and F. Yu, An improved Hirota bilinear method and new application for a nonlocal integrable complex modified Korteweg-de Vries (MKdV) equation. \textit{Physics Letters A} \textbf{383}, 1578-1582 (2019).

\bibitem{Bai}
Y.-S. Bai, L.-N. Zheng, W.-X. Ma and Y.-S. Yun, Hirota bilinear approach to multi-component nonlocal nonlinear Schrödinger equations. \textit{Mathematics} \textbf{12}(16), 2594 (2024).

\bibitem{Carstea}
A. S. Carstea, Extension of the bilinear formalism to supersymmetric KdV-type equations. \textit{Nonlinearity} \textbf{13}, 1645 (2000).

\bibitem{ZhangLiu}
M. Zhang, Q. Liu, Y. Shen \textit{et al.}, Bilinear approach to N=2 supersymmetric KdV equations. \textit{Sci. China Ser. A-Math.} \textbf{52}, 1973-1981 (2009).

\bibitem{Grammaticos}
B. Grammaticos, A. Ramani and A. S. Carstea, Bilinearization and soliton solutions of the N=1 supersymmetric sine-Gordon equation. \textit{J. Phys. A: Math. Gen.} \textbf{34}, 4881 (2001) .

\bibitem{DelisleMas}
L. Delisle and M. Mosaddeghi  Classical and SUSY solutions of the Boiti-Leon-Manna-Pempinelli equation. \textit{J. Phys. A: Math. Theor.} \textbf{46}, 115203 (2013).

\bibitem{Ma} W.-X. Ma,  Lump solutions to the Kadomtsev-Petviashvili equation. \textit{Physics Letters A} \textbf{379}, 1975-1978 (2015).

\bibitem{Yue}
Y. Yue, L. Huang and Y. Chen, N-solitons, breathers, lumps and rogue wave solutions to (3+1)-dimensional nonlinear evolution equation. \textit{Computers and Mathematics with Applications} \textbf{75}, 2538-2548 (2018).

\bibitem{Clarkson}
P. A. Clarkson, Remarks on the Yablonskii-Vorob'ev polynomials. \textit{Physics Letters A} \textbf{319}, 137-144 (2003).

\bibitem{MaBell}
W.-X. Ma, Bilinear equations, Bell polynomials and linear superposition principle. \textit{J. Phys.: Conf. Ser.} \textbf{411}, 012021 (2013).

\bibitem{MaTril}
W.-X. Ma, Trilinear equations, Bell polynomials, and resonant solutions. \textit{Frontiers of Mathematics in China} \textbf{8}, 1139-1156 (2013).

\bibitem{MaSub}
W.-X. Ma, Y. Zhang, Y. Tand and J. Tu,  Hirota bilinear equations with linear subspaces of solutions. \textit{Applied Mathematics and Computation} \textbf{218}, 7174-7183 (2012).

\bibitem{DelisleOCNMP}
L. Delisle and A. Jaouadi, A vector bilinear framework for soliton dynamics in coupled modified KdV systems, \textit{Open Commun. Nonlinear Math. Phys.}, Special Issue: Hietarinta, ocnmp:18025, 35-52, (2026).

\bibitem{Hirota1}
R. Hirota, Molecule solutions of coupled modified KdV equations. \textit{J. Phys. Soc. Jpn.} \textbf{66}, 2530 (1997).

\bibitem{Iwao}
M. Iwao and R. Hirota, Soliton solutions of a coupled modified KdV equations. \textit{J. Phys. Soc. Jpn.} \textbf{66}, 577-588 (1997).



\bibitem{Ohta}
Y.~Ohta \emph{et al.},
``General $N$-Dark-Dark Solitons in the Coupled Nonlinear Schr\"odinger Equations,''
\emph{Studies in Applied Mathematics} \textbf{127}, 345--371 (2011).

\bibitem{Jiang}
Y.~Jiang \emph{et al.},
``Soliton interactions and complexes for coupled nonlinear Schr\"odinger equations,''
\emph{Phys. Rev. E} \textbf{85}, 036605 (2012).


\bibitem{Vijaya}
M.~Vijayajayanthi \emph{et al.},
``Bright-dark solitons and their collisions in mixed $N$-coupled nonlinear Schr\"odinger equations,''
\emph{Phys. Rev. A} \textbf{77}, 013820 (2008).

\bibitem{Radha}
R.~Radhakrishnan and K.~Aravinthan,
``A dark-bright optical soliton solution to the coupled nonlinear Schr\"odinger equation,''
\emph{J. Phys. A: Math. Theor.} \textbf{40}, 13023--13030 (2007).

\bibitem{Rad}
R.~Radhakrishnan and M.~Lakshmanan,
``Bright and dark solitons to coupled nonlinear Schr\"odinger equations,''
\emph{J. Phys. A: Math. Gen.} \textbf{28}, 2683--2692 (1995).

\bibitem{Kanna1}
T.~Kanna and M.~Lakshmanan,
``Exact soliton solutions, shape changing collisions, and partially coherent solitons in coupled nonlinear Schr\"odinger equations,''
\emph{Phys. Rev. Lett.} \textbf{86}, 5043 (2001).

\bibitem{MaoZhao}
N.~Mao and L.-C.~Zhao,
``Exact analytical soliton solutions of $N$-component coupled nonlinear Schr\"odinger equations with arbitrary nonlinear parameters,''
\emph{Phys. Rev. E} \textbf{106}, 064206 (2022).
\bibitem{Trippenbach2000}
M.~Trippenbach, K.~G\'oral, K.~Rz\k{a}\.zewski, B.~A.~Malomed, and Y.~B.~Band,
``Structure of binary Bose--Einstein condensates,''
\emph{J. Phys. B: At. Mol. Opt. Phys.} \textbf{33}, 4017--4031 (2000).

\bibitem{Kevrekidis2016}
P.~G.~Kevrekidis and D.~J.~Frantzeskakis,
``Solitons in coupled nonlinear Schr\"odinger models: A survey of recent developments,''
\emph{Review in Physics} \textbf{1}, 140--153 (2016).


\bibitem{Mollenauer}
L.~F.~Mollenauer \emph{et al.},
``Experimental observation of picosecond pulse narrowing and solitons in optical fibers,''
\emph{Phys. Rev. Lett.} \textbf{45}, 1095 (1980).

\bibitem{Hasegawa}
A.~Hasegawa \emph{et al.},
``Transmission of stationary nonlinear optical pulses in dispersive dielectric fibers. I. Anomalous dispersion,''
\emph{Appl. Phys. Lett.} \textbf{23}, 142 (1973).

\bibitem{LiuPu} X. Liu \emph{et al.}  Formation and transformation of vector solitons in two-species Bose-Einstein condensates with tunable interaction. \textit{Phys. Rev. A} \textbf{79}, 013423 (2009).

\bibitem{Manakov}
S.~V.~Manakov,
``On the theory of two-dimensional stationary self-focusing of electromagnetic waves,''
\emph{Soviet Physics-JETP} \textbf{38}, 248 (1974).

\bibitem{Zhang}
X.-F.~Zhang \emph{et al.},
``Vector solitons in two-component Bose--Einstein condensates with tunable interactions and harmonic potential,''
\emph{Phys. Rev. A} \textbf{79}, 033630 (2009).

\bibitem{Timmermans}
E.~Timmermans,
``Phase separation of Bose--Einstein condensates,''
\emph{Phys. Rev. Lett.} \textbf{81}, 5718 (1998).

\bibitem{Ohberg2001}
P.~\"Ohberg and L.~Santos,
``Dark solitons in a two-component Bose--Einstein condensate,''
\emph{Phys. Rev. Lett.} \textbf{86}, 2918--2921 (2001).


\bibitem{Morera2018}
I.~Morera, G.~Theocharis, P.~Schmelcher, P.~G.~Kevrekidis, and D.~J.~Frantzeskakis,
``Dark--dark-soliton dynamics in two-density-coupled Bose--Einstein condensates,''
\emph{Phys. Rev. A} \textbf{97}, 043621 (2018).

\bibitem{Kanna}
T.~Kanna \emph{et al.},
``Non-autonomous bright-dark solitons and Rabi oscillations in multi-components Bose--Einstein condensates,''
\emph{J. Phys. A: Math. Theor.} \textbf{46}, 475201 (2013).

\bibitem{Busch}
Th.~Busch and J.~R.~Anglin,
``Dark-Bright Solitons in Inhomogeneous Bose--Einstein Condensates,''
\emph{Phys. Rev. Lett.} \textbf{87}, 010401 (2001).











\bibitem{Perez2005} V. M. P\'erez-Garc\`ia and J. B. Beitia Symbiotic solitons in heteronuclear multicomponent Bose-Einstein condensates. \textit{Phys. Rev. A} \textbf{72}, 033620  (2005). 













\bibitem{DingPRE}
C.-C. Ding, Q. Zhou and B. A. Malomed, Beating stripe solitons arising from helicoidal spin-orbit coupling in Bose-Eintein condensates, \textit{Phys. Rev. E} \textbf{111}, 044203 (2025).



\end{thebibliography}
\end{document}